\documentclass{tcibook}
\usepackage{fancyhea}
\usepackage{work}
\usepackage{bm}       

\usepackage{graphicx}
\usepackage{multirow}
\usepackage{lineno}
\usepackage{threeparttable}
\usepackage[normalem]{ulem}
\usepackage{color}
\usepackage[colorlinks = true,
            linkcolor = blue,
            urlcolor  = blue,
            citecolor = blue,
            anchorcolor = blue]{hyperref}

\def\beq{\begin{equation}}
\def\eeq#1{\label{#1}\end{equation}}
\def\eeqn{\end{equation}}

\newenvironment{Eqnarray}%
   {\arraycolsep 0.14em\begin{eqnarray}}{\end{eqnarray}}
\def\beqa{\begin{Eqnarray}}
\def\eeqa#1{\label{#1}\end{Eqnarray}}
\def\eeqan{\end{Eqnarray}}

\let\bar=\overbar

\def\lsim{\mathrel{\raise.3ex\hbox{$<$\kern-.75em\lower1ex\hbox{$\sim$}}}}
\def\gsim{\mathrel{\raise.3ex\hbox{$>$\kern-.75em\lower1ex\hbox{$\sim$}}}}

\def\del{\partial}
\def\Dslash{\not{\hbox{\kern-4pt $D$}}}
\def\dslash{\not{\hbox{\kern-2pt $\del$}}}
\def\pslash{\not{\hbox{\kern-2pt $p$}}}
\def\ETmiss{\not{\hbox{\kern-4pt $E$}}_T}

\def\Dlr{\mathrel{\raise1.5ex\hbox{$\leftrightarrow$\kern-1em\lower1.5ex\hbox{$D$}}}}

\def\MSB{{\bar{M \kern -2pt S}}}
\def\msb{{\bar{\scriptsize M \kern -1pt S}}}

\def\drb{{\bar{\scriptsize D \kern -1pt R}}}

\def\authorlist#1#2{
    \vskip 0.4in
\begin{center}\begin{large} {\bf  #1 } \end{large}
    \vskip 0.2in
              #2
     \vskip 0.2in
   \end{center}
}
\pdfoutput=1

\begin{document}


\pagenumbering{roman}
\pdfoutput=1
\parindent=0pt
\parskip=8pt
\setlength{\evensidemargin}{0pt}
\setlength{\oddsidemargin}{0pt}
\setlength{\marginparsep}{0.0in}
\setlength{\marginparwidth}{0.0in}
\marginparpush=0pt


\pagenumbering{arabic}

\renewcommand{\chapname}{chap:intro_}
\renewcommand{\chapterdir}{.}
\renewcommand{\arraystretch}{1.25}
\addtolength{\arraycolsep}{-3pt}

\setcounter{chapter}{6} 


\chapter{Accelerator Technology – Magnets}

\newcommand*{\affaddr}[1]{#1} 
\newcommand*{\affmark}[1][*]{\textsuperscript{#1}}

\vspace{0.3in}

\begin{center}\begin{Large}  Topical Group Conveners   \end{Large}\end{center} 
\vspace{-0.3in}
\authorlist{
S. Izquierdo Bermudez\affmark[6], 
G. Sabbi\affmark[2], 
A.V. Zlobin\affmark[1]}
 { }
 
\begin{center}\begin{Large}  White Paper Contributors   \end{Large}\end{center} 
\vspace{-0.4in}
\authorlist{
Y. Alexahin\affmark[1], 
G. Ambrosio\affmark[1],
G. Apollinari\affmark[1],
M. Baldini\affmark[1],
R. Carcagno\affmark[1],
E. Barzi\affmark[1],
C. Boffo\affmark[1],
G. Chlachidze\affmark[1],
B. Claypool\affmark[1],
J. DiMarco\affmark[1],
S. Feher\affmark[1],
E. Gianfelice-Wendt\affmark[1], 
S. Hays\affmark[1],
D. Hoang\affmark[1],
V. Kapin\affmark[1],
V. Kashikhin\affmark[1],
V. V. Kashikhin\affmark[1],
S. Krave\affmark[1],
M. Kufer\affmark[1],
J. Lee\affmark[1],
V. Lombardo\affmark[1],
V. Marinozzi\affmark[1],
N. Mokhov\affmark[1],
F. Nobrega\affmark[1],
I. Novitski\affmark[1],
X. Peng\affmark[1],
H. Piekarz\affmark[1],
V. Shiltsev\affmark[1],
S. Stoynev\affmark[1],
T. Strauss\affmark[1],
S. Striganov\affmark[1], 
M. Suarez\affmark[1],
N. Tran\affmark[1],
I. Tropin\affmark[1],
M. Turenne\affmark[1],
D. Turrioni\affmark[1],
G. Velev\affmark[1],
X. Xu\affmark[1],
A.V. Zlobin\affmark[1],
D. Arbelaez\affmark[2],
A. Baskys\affmark[2],
C. Bird\affmark[2],
J. Boerme\affmark[2],
L. Brouwer\affmark[2],
S. Caspi\affmark[2],
D. Dietderich\affmark[2],
L. Garcia Fajardo\affmark[2],
J. Rudeiros Fernandez\affmark[2],
P. Ferracin\affmark[2],
S. Gourlay\affmark[2],
A. Hafalia\affmark[2],
H. Higley\affmark[2],
M. Juchno\affmark[2],
G. Lee\affmark[2],
L. Luo\affmark[2],
M. Marchevsky\affmark[2],
C. Messe\affmark[2],
C. Myers\affmark[2],
M. Naus\affmark[2],
I. Pong\affmark[2],
S. Prestemon\affmark[2],
G. Sabbi\affmark[2],
T. Shen\affmark[2],
R. Teyber\affmark[2],
M. Turqueti\affmark[2],
G. Vallone\affmark[2],
S. Viarengo\affmark[2],
L. Wang\affmark[2],
X. Wang\affmark[2],
S. Yin\affmark[2],
K. Amm\affmark[3],
M. Anerella\affmark[3],
J. Avronsart\affmark[3],
A. Ben Yahia\affmark[3],
J. Cozzolino\affmark[3],
R. Gupta\affmark[3],
P. Joshi\affmark[3],
M. Kumar\affmark[3],
F. Kurian\affmark[3],
J. Muratore\affmark[3],
M. Palmer\affmark[3],
B. Parker\affmark[3],
C. Runyan\affmark[3],
W. Sampson\affmark[3],
J. Schmalzle\affmark[3],
S. Balachandran\affmark[4],
S. Barua\affmark[4],
E. Bosque\affmark[4],
N. Cheggour\affmark[4],
L. Cooley\affmark[4],
D. Davis\affmark[4],
L. English\affmark[4],
E. Hellstrom\affmark[4],
I. Hossain\affmark[4],
J. Jiang\affmark[4],
F. Kametani\affmark[4],
D. Larbalestier\affmark[4],
P. Lee\affmark[4],
C. Tarantini\affmark[4],
U. Trociewitz\affmark[4],
M. Abe\affmark[5],
M. Iio\affmark[5],
Y. Makida\affmark[5],
T. Nakamoto\affmark[5],
T. Okamura\affmark[5],
T. Ogitsu\affmark[5],
K. Sasaki\affmark[5],
M. Sugano\affmark[5],
N. Sumi\affmark[5],
K. Suzuki\affmark[5],
A. Yamamoto\affmark[5,6],
M. Yoshida\affmark[5], 
D. Aguglia\affmark[6],
A. Ballarino\affmark[6],
F. Boattini\affmark[6],
L. Bottura\affmark[6],
B. Cure\affmark[6],
N. Deelen\affmark[6],
A. Dudarev\affmark[6],
R. Losito\affmark[6],
M. Mentink\affmark[6],
D. Schulte\affmark[6],
B. Baudouy\affmark[7],
P. Fazilleau\affmark[7],
L. Quettier\affmark[7],
E. Rochepault\affmark[7],
P. Vedrine\affmark[7],
L. García-Tabarés\affmark[8],
J. Munilla\affmark[8], 
F. Toral\affmark[8],
B. Auchmann\affmark[9]
T. Arndt\affmark[10],
M. Noe\affmark[10],
E. De Matteis\affmark[11],
S. Mariotto\affmark[11,24],
M. Prioli\affmark[11],
L. Rossi\affmark[11,24],
M. Sorbi\affmark[11,24],
M. Statera\affmark[11],
R. Valente\affmark[11],
A. Bersani\affmark[12],
B. Caiffi\affmark[12],
S. Farinon\affmark[12],
R. Musenich\affmark[12],
X. Chaud\affmark[13],
F. Debray\affmark[13],
B. Shepherd\affmark[14],
A. Lankford\affmark[15],
M. Jewell\affmark[16],
J. Minervini\affmark[17]
J. Rochester\affmark[19],
M. Sumption\affmark[19]
C. Senatore\affmark[20],
J. Bear\affmark[21],
M. Dam\affmark[22],
H. De Gersem\affmark[22],
A. Kario\affmark[23],
H. Ten Kate\affmark[23],
M. Breschi\affmark[25],
Y. Yang\affmark[26],
Q. Xu\affmark[27],
S. Awaji\affmark[28],
N. Amemiya\affmark[29]
H. Iinuma\affmark[30],
S. Kahn\affmark[31], 
R. Scanlan\affmark[31], 
R. Weggel\affmark[31], 
E. Willen\affmark[31],
D. van der Laan\affmark[32],
J. Weiss\affmark[32]
P. McIntyre\affmark[33,34]
}
\affaddr{\affmark[1]FNAL, Batavia, Illinois, United States}\\
\affaddr{\affmark[2]LBNL, Berkeley, California, United States}\\
\affaddr{\affmark[3]BNL, Upton, New York, United States}\\
\affaddr{\affmark[4]FSU/NHMFL, Tallahasse, Florida, United States}\\
\affaddr{\affmark[5]KEK, Tsukuba, Ibaraki, Japan}\\
\affaddr{\affmark[6]CERN, Geneva, Switzerland}\\
\affaddr{\affmark[7]CEA, Saclay, France}\\
\affaddr{\affmark[8]CIEMAT, Madrid, Spain}\\
\affaddr{\affmark[9]PSI, Villigen, Switzerland}\\
\affaddr{\affmark[10]KIT/ITEP, Karlsruhe, Germany}\\
\affaddr{\affmark[11]INFN/LASA, Milan, Italy}\\
\affaddr{\affmark[12]INFN Sezione di Genova, Italy}\\
\affaddr{\affmark[13]LNCMI, Toulouse, France}\\
\affaddr{\affmark[14]STFC/RAL, Harwell Campus, United Kingdom}\\
\affaddr{\affmark[15]UC Irvine, California, United States}\\
\affaddr{\affmark[16]University of Wisconsin-Eau Claire, Winsconsin, United States}\\
\affaddr{\affmark[17]MIT, Cambridge, Massachusetts, United States}\\
\affaddr{\affmark[18]Politecnico di Torino, Torino, Italy}\\
\affaddr{\affmark[19]Ohio State University, Columbus, Ohio, United States}\\
\affaddr{\affmark[20]University of Geneva, Switzerland}\\
\affaddr{\affmark[21]LNCMI, Grenoble, France}\\
\affaddr{\affmark[22]Technical University of Darmstadt, Germany}\\
\affaddr{\affmark[23]University of Twente, The Netherlands}\\
\affaddr{\affmark[24]Universita' di Milano, Italy}\\
\affaddr{\affmark[25]Universita' di Bologna, Italy}\\
\affaddr{\affmark[26]University of Southampton, United Kingdom}\\
\affaddr{\affmark[27]IHEP-CAS, Beijing, China}\\
\affaddr{\affmark[28]Tohoku University, Japan}\\
\affaddr{\affmark[29]Kyoto University, Japan}\\
\affaddr{\affmark[30]Ibaraki University, Japan}\\
\affaddr{\affmark[31]Particle Beam Lasers Inc., Texas, United States}\\
\affaddr{\affmark[32]Advanced Conductor Technologies LLC, Colorado, United States}\\
\affaddr{\affmark[33]TAMU, Texas, United States}\\
\affaddr{\affmark[34]Accelerator Technology Corp., Texas, United States}\\
\\

\newpage

\begin{center}\begin{Large} \textbf{ Executive Summary}   \end{Large}\end{center} 

\vspace{0.1in}

The performance of high energy colliders is strongly dependent on their magnet system: arc dipoles for energy reach, and interaction region quadrupoles for luminosity. As the HEP community explores its best options to enable future discoveries, magnet technology considerations are essential to make informed decisions on the feasibility and cost of achieving the physics goals.

\vspace{0.2in}
\underline{Conductor technologies}
\vspace{0.1in}

Three classes of superconducting materials can be considered for magnet applications: Nb-Ti up to 8 T, Nb\textsubscript{3}Sn up to 16 T, and High Temperature Superconductors (HTS) beyond 16 T. These materials can be also combined in hybrid designs to optimize magnet cost and performance. 

Excellent mechanical and electrical properties of multifilamentary Nb-Ti have made it the conductor of choice in superconducting accelerator and detector magnets starting from the Tevatron. It allows simple fabrication methods for wires and cables, and does not suffer from performance degradation under mechanical stress. However, it has limited operational range in terms of both field ($<$ 8 T) and temperature ($<$ 5 K)   

Nb\textsubscript{3}Sn wires are available from industry in long lengths with uniform properties and carry currents comparable to Nb-Ti wires of the same size at more than twice the field. The upgraded Interaction Region Quadrupoles of the High Luminosity LHC will be the first Nb\textsubscript{3}Sn application in a high energy collider. Further improvements of critical current density (J\textsubscript{c}) and specific heat (C\textsubscript{p}) are being demonstrated in recent R\&D wires. A reduction of the effective filament diameter is also required for future arc dipoles, but has proven difficult to attain while preserving high J\textsubscript{c} and controlling cost. In addition, maintaining industrial production capabilities for the wires of interest to the HEP community is becoming a challenge, in part due to lack of large orders after completion of the ITER production. 

The most promising HTS conductors are Bi2212 and REBCO. Bi2212 is available in round isotropic wires and can generally follow the design and fabrication approach developed for Nb\textsubscript{3}Sn. A significant increase of J\textsubscript{c} has been obtained in recent years with improved powders and high-pressure heat treatment. However, Bi2212 technology is significantly more challenging than Nb\textsubscript{3}Sn due to higher strain sensitivity, high required reaction temperature in an oxygen-rich environment, and chemical compatibility with the cable insulation and coil structural materials during heat treatment. Additional challenges derive from the high cost and limited industrial suppliers. 

REBCO is produced by deposition on thin tapes that include layers of metals, oxides and ceramics for crystal plane alignment, mechanical strength and electrical stability. Steady improvements in engineering current density, uniformity and piece length have been obtained. The main challenge for application of these conductors to accelerator magnets is the development of cables capable of carrying high currents while retaining sufficient flexibility to be wound in small-aperture coils with minimal degradation. Additional challenges are represented by large magnetization effects; electromagnetic anisotropy requiring good alignment of the tape to the field direction for optimal current carrying capability; and slow quench propagation velocities requiring new approaches to quench detection and protection. 

A new class of Iron Based Superconductors has attracted  interest by the magnet community in recent years. They show significant potential for high Jc at high fields, and lower cost. However, fundamental questions regarding intrinsic Jc mechanisms, possibility of multifilament wires, and production capabilities need to be addressed before they can be considered for large scale applications. 

\vspace{0.2in}
\underline{Accelerator magnets}
\vspace{0.1in}

High-field dipoles and quadrupoles are the enabling technology for hadron colliders, and critical for muon colliders. All the key coil layouts including shell (cosine-theta), block, common coil, canted cosine-theta are being explored and compared in terms of their performance and cost. The main goals for the next decade are:
\begin{itemize}
\item Development of Nb\textsubscript{3}Sn magnet technology for collider-scale production in the 12-14 T range through robust design, cost reduction, and preparations for industrialization. 
\item Development and study of coil stress management structures for Nb\textsubscript{3}Sn magnets in the field range of 14-16 T.
\item Development of insert coils based on Bi2212 and REBCO conductor, test in self field and background field. 
\item Development and test of short models using hybrid HTS/LTS approach to increase field toward 20 T.
\item Exploring all HTS magnets for higher temperature operation (e.g. to reduce power consumption). Leverage the HTS technology development for applications such as fusion power, offshore wind, aviation and hydrogen economy.
\item Understanding strain dependence characteristics, including test in high background field, and developing new designs and fabrication approaches to minimize strain degradation, since all advanced superconductors are brittle and strain sensitive.
\end{itemize}
Two additional magnet categories have been identified as being of particular interest in connection with the muon collider development. They include a) fast ramping magnets with a ramp rate up to 400 T/m, while increasing the field amplitude from 0.5 to ±2+ T, and b) high-field solenoids for muon cooling with 50 mm aperture and 40+ T bore field. These magnets require HTS which provides synergy with D and Q accelerator magnet technologies.

\vspace{0.1in}
\underline{Detector magnets }
\vspace{0.1in}

Superconducting detector magnets are a key component of particle physics experiments. Over the past four decades, the design and fabrication of detector magnets has made significant progress thanks to the introduction and demonstration of specific technologies such as: high current, high strength, Al-stabilized superconducting cables; coil winding techniques and coil support structures; improved transparency to collision products using indirect magnet cooling; etc. Detector magnets for future colliders will be based on these technologies but further improvements are required to cope with larger size, radiation loads and physics performance requirements. Due to large intervals between projects, there are concerns about maintaining expertise and industrial capabilities. A sustained effort is needed to preserve and further develop the technologies for application in future detector magnets, in particular Al-stabilized SC cable with both higher strength and high RRR. In addition, the use of HTS should be considered to reduce cryogenics power consumption of detector magnets. A collaborative framework between institutes and industries is critical.

Finally, there is significant interest in the development of magnets for axion searches. Here there are synergies with the development of accelerator dipoles, and even the possibility of adapting large aperture prototypes (e.g. muon collider arc dipoles) for use in axion detectors. 

\vspace{0.2in}
\underline{Performance optimization, diagnostics and modelling} 
\vspace{0.1in}

Several potential techniques have emerged that may improve magnet performance. They include high-Cp insulation that promises enabling stable operation against perturbations; composite magnet components for both Nb\textsubscript{3}Sn and HTS magnets; thermal management of high field magnets including internal heat transfer, heat transfer to coolant, and heat transfer to cryo-plant. 

New diagnostics and testing techniques are providing important insight into magnet behavior and provide critical feedback to conductor and magnet designers. Examples include fiber optics, acoustics, cold-electronics, and fast controls. Advances in modelling provide more detailed insight into stress and strain states in magnets, including full 3D effects and complex nonlinear material and interface behavior.

\vspace{0.1in}
\underline{Facilities} 
\vspace{0.1in}

The development of novel SC magnet technology at the high-field frontier requires specialized fabrication and test infrastructure. The necessary investment is considerable, so the effort should extend across national and international labs. In the U.S. the magnet R\&D infrastructure exists at the three National Laboratories (BNL, FNAL, LBNL) as well as at NHMFL and TAMU. Some capabilities are unique including long coil winding-reaction-impregnation facilities; long magnet assembly and test facilities at BNL and FNAL; 60-strand cabling machine at LBNL; 10 T common coil facility for cable and insert magnet testing at BNL; ovens for HTS coil reaction at high pressure at NHMFL. At present, FNAL in collaboration with LBNL is building a new high-field cable testing facility that will serve both the HEP and FES Programs for testing high current cables in a 15 T field. 

It is critical that continued adequate support is provided to operate, maintain, and upgrade the existing U.S. accelerator magnet fabrication and test facilities. At the same time, collaborations among U.S. Laboratories as well as with European and Japanese laboratories, universities and industry are fully exploited to enhance capabilities and reduce the cost of material and magnet development and test.

\vspace{0.1in}
\underline{Magnet R\&D programs} 
\vspace{0.1in}

The U.S. Magnet Development Program (US-MDP) focuses on fundamental accelerator magnet R\&D providing key magnet technologies which benefit all future accelerators. It has emerged as a successful ongoing national program with short- and long-term goals and plans, efficient management structure, established staff, and facilities. It is also well-recognized internationally thanks to active participation in international magnet R\&D activities. 

Discussions and initiatives emerging in the Snowmass’21 process provide a strong motivation to enhance the scope and significant increase in resources of accelerator magnet technology. In particular, US-MDP scope and resources need to be enhanced to address general magnet R\&D needs for HEP, including magnet design and short model development for specific accelerator applications, e.g. muon collider. Additional investments in critical technology scale up and cost reduction efforts and preparation for industrialization were proposed during the Snowmass process, such as the Leading-Edge technology And Feasibility directed (LEAF) program. These efforts will inform the design of future accelerators by providing a sound basis for facility specifications and costing. 

\thispagestyle{empty}

\tableofcontents

\newpage

\section{Introduction}

The Snowmass community exercise started in April 2020 to identify and document a scientific vision for the future of particle physics in the US and international partners. The AF7-Magnets working group was charged to a) address the potential contributions of magnet technology to future HEP facilities, b) evaluate the R\&D required to enable these opportunities, c) estimate the time and cost scales of these efforts, and d) assess the needs for associated fabrication infrastructure and test facilities.

Following a series of meetings to introduce the program goals and process, Letters of Interest (LoI) were solicited to get input from the community. By October 2020, the AF7-Magnets topical group had received a total of 50 LoI which were subdivided in 6 groups: Regional Plans (8 LoI), Conductor (16 LoI), High Field Magnet Development (9 LoI), Magnet Technologies (8 LoI), Detector Magnets (4 LoI) and Special Magnets (5 LoI). Because of the coronavirus pandemic, a pause was decided in January 2021. The process resumed in fall 2021 with a call for White Papers describing technical progress and opportunities in specific areas, as well as recommended plans and priorities for the next decade. In March 2022, the Magnet topical group received 21 White Papers (WPs) ~\cite{A7-1} -~\cite{A7-21} which were used as a basis for this summary report.  

This report addresses the working group charge, summarizes the status of accelerator and detector magnet technologies, and discuss ideas and plans to push this key area of the US and international HEP to new horizons.
\section{Magnet needs for future colliders}
HEP colliders depend on accelerator magnets to guide and focus the particle beams, and on detector magnets to analyze the momentum and polarity of collision products. In this section, we briefly review the specific requirements for different machines.

Hadron colliders presently under consideration include: the very large circumference, 500 TeV “Collider-in-Sea” based on cost-efficient 4 T Nb-Ti magnets; the 100 km ring Future Circular Collider (FCC-hh) based on 16 T Nb\textsubscript{3}Sn dipoles, the practical limit for this material; the 50-100 km ring Super proton-proton Collider (SPPC) based on 12-24 T dipoles to achieve 75-150 GeV center-of-mass energy; and two smaller circumference machines, High Energy Large Hadron Collider (HE-LHC) and the Fermilab site-filler, where existing site/infrastructure considerations limit the ring circumference, requiring dipole fields within the range of 16-26 T to achieve sufficient beam energy. 

Muon colliders require very strong solenoids for efficient muon production and cooling. Additional challenges to the magnet system are posed by radiation load at the muon production target. The short muon lifetime motivates novel magnet and optics schemes to enable very rapid acceleration from production target to the storage ring where collisions occur. Examples include very-fast ramping magnets in a synchronous multi-pass configuration, for example using high-temperature superconductors in novel configurations, and/or large momentum acceptance magnet optics, such as fixed-field alternating gradient (FFAG) designs that enable the beam acceleration via multiple passes with fixed magnetic field. Finally, since even at relativistic velocities the lifetime of the muon is limited, luminosity considerations motivate the most compact storage ring possible, requiring the highest field dipoles possible to minimize the ring circumference, while coping with severe radiation load from muon decay.

IR magnets are also of critical importance for the Collider physics reach. In addition to hadron and muon colliders, this applies to electron-positron colliders, both linear and circular. As already demonstrated by past projects from LEP to HL-LHC, special magnet technology can be deployed in the IR, different from the rest of the machine. The relatively small scale of these systems makes it possible to adopt solutions that push the state of the art, and with potentially higher unit cost, in order to achieve the highest possible Luminosity. 

\section{Overview of the HEP accelerator magnet programs}
The magnet system is a critical component and major cost driver for future HEP colliders. Magnet R\&D programs are underway to take advantage of new developments in superconducting materials, achieve higher efficiency and simplify fabrication while preserving accelerator-class field quality. This section provides a summary of the major ongoing and planned activities worldwide.
\subsection{United States}
\subsubsection{HL-LHC Accelerator Upgrade Project}
The US government is making an investment of more than 750 M\$ in the upgrade of the LHC to achieve High Luminosities necessary to fully exploit the HEP frontier at the LHC energies. These investments will support the construction of upgraded CMS and ATLAS detectors, and new Interaction Regions (IRs) that will enable a 10-fold increase of the luminosity delivered to the detectors. The upgraded machine, called HL-LHC, is presently in its construction phase. The US is contributing to the HL-LHC accelerator through the DOE Accelerator Upgrade Project (AUP) ~\cite{A7-23}, executed by a collaboration of National Laboratories (FNAL, BNL, LBNL). US-built IR quadrupoles and crab cavities will be deployed at CERN for installation and commissioning of HL-LHC in the 2026-2029 period. 

\subsubsection{US Magnet Development Program}

Following the 2013-2014 Snowmass and P5 process, DOE-OHEP initiated the US Magnet Development Program (MDP), a general R\&D program that integrates DOE magnet laboratories and University programs under a common collaboration ~\cite{A7-1}. The program has oversight through a longstanding Technical Advisory Committee, composed of internationally recognized experts, as well as a Steering Council composed of laboratory leadership from the MDP collaborating institutions.
MDP advances fundamental aspects of magnet technology with potential benefits to a broad range of collider applications, including hadron colliders and muon colliders. Over the years, the MDP collaboration has expanded through synergies with the NSF-funded NHMFL; with other DOE offices, in particular DOE-OFES; and with industry, for example through the SBIR program. Furthermore, international collaborations exist on a range of technical topics. The program goals are articulated in four main elements:
\begin{itemize}
\item Explore the performance limits of Nb\textsubscript{3}Sn accelerator magnets, with a focus on minimizing the required operating margin and reducing training;
\item Develop and demonstrate an HTS accelerator magnet with a self-field of 5 T or greater, compatible with operation in a hybrid HTS/LTS magnet for fields beyond 16 T;
\item Investigate fundamental aspects of magnet design and technology that can lead to substantial performance improvements and magnet cost reduction;
\item Pursue Nb\textsubscript{3}Sn and HTS conductor R\&D with clear targets to increase performance, understand present performance limits, and reduce the cost of accelerator magnets.
\end{itemize}
In the magnet design and fabrication area, a major theme is the development of methods to control the conductor strain, in particular through the use of “stress-managed” coil layouts and support structures. Two main approaches are being explored, the “Canted Cosine Theta” ~\cite{A7-24} and the Stress-Management Cosine-Theta ~\cite{A7-25}. While initially developed on Nb\textsubscript{3}Sn, these principles are being applied to other stress/strain sensitive superconductors, such as High Temperature Superconductors (HTS).

In the area of high field solenoids for muon production and for muon beam cooling MDP has strong synergies with ongoing R\&D at the National High Magnetic Field Laboratory (NHMFL), as well as advances in REBCO magnet technology by DOE and by private industry toward compact, very high field magnets for fusion reactors.

The development of advanced superconductors specifically optimized for accelerator magnet applications is at the base of the MDP effort.  Over the years the US has built a strong technical and industrial leadership in this area, with significant contributions from projects like the Tevatron and SSC, as well as the HEP-funded Conductor Development Program. MDP is now coordinating these efforts, working closely with the new Accelerator R\&D and Production Office (ARDAP) to develop a strategic plan that can strengthen US industry in this arena and support HEP's long term needs for conductor performance and cost-effective conductor production.

Supporting technologies critical to design optimization and performance assessment are also of major interest. These include advanced modelling capabilities, novel diagnostics, and new concepts that can impact the rate at which accelerator magnets “train" up to full field ~\cite{A7-15} -~\cite{A7-18}. Advances in these areas provide lasting benefit for HEP and are highly valued by the broader superconducting magnet community, both in research and industry.

A long-range magnet R\&D program, designed to advance magnet technology while fully leveraging the broader community's strengths, is vital to HEP and the future of particle physics. The MDP experience indicates that a 15M\$ per year investment is needed to pursue fundamental magnet R\&D and to provide a foundation for a directed program aimed at project-specific design demonstration and production readiness. Additional 3M\$ per year in conductor procurement from industry is required to support the magnet development needs.

\subsubsection{Directed R\&D}
The development and demonstration of maturity of advanced magnet technology for application in the present upgrades to the LHC (High Luminosity LHC Upgrade, HL-LHC) was made possible by a $\sim$ 15 years-long US national program of directed R\&D (called LHC Accelerator Research Program, LARP) working in combination with generic and complementary R\&D efforts (Conductor Development Program, General Accelerator R\&D GARD, university programs, etc.).
A similar Leading-Edge technology And Feasibility-directed (LEAF) Program is proposed ~\cite{A7-2} to achieve readiness for a future collider decision on the timescale of the next decade. As for its predecessor, the LEAF Program would rely on, and be synergetic with, generic R\&D efforts presently covered in the US by the Magnet Development Program (MDP), the Conductor Procurement and R\&D (CPRD) Program and other activities in the Office of HEP supported by Early Career Awards (ECA) or Lab Directed R\&D (LDRD) funds. Where possible, ties to synergetic efforts in other Offices of DOE or NSF are highlighted and suggested as wider Collaborative efforts on the National scale. International efforts are also mentioned as potential partners in the LEAF Program.
The LEAF Program will concentrate on demonstrating the feasibility of magnets for muon colliders as well as next generation high-energy hadron colliders, pursuing, where necessary and warranted by the nature of the application, the transition from R\&D models to long models/prototypes. The LEAF Program will naturally drive accelerator-quality and experiment interface design considerations. LEAF will also concentrate, where necessary, on cost reduction and/or industrialization steps.
The LEAF Program is foreseen to be a decade-long effort starting around 2024-2025 to be concluded on the timescale of 2034-2035. Based on the experience of the proponents, the appropriate funding level for the LEAF Program should be 25-30M\$/year across the spectrum of participants (US National Laboratories and Universities).
\subsection{Europe}
Over the past several decades, a strong magnet R\&D effort was pursued in Europe through the EU Framework Programs including: EU FP6 Coordinated Accelerator Research in Europe (CARE); EU FP7 European Coordination for Accelerator Research and Development (EuCARD); EU FP7 Enhanced European Coordination for Accelerator Research and Development (EuCARD2); EU FP7 Accelerator Research and Innovation for European Science and Society (ARIES); and current work such as HL-LHC, EU H2020 Innovation Fostering in Accelerator Science and Technology (I-FAST), and CERN-HFM. 

Looking at the future, in 2021 the European Laboratory Directors Group (LDG) was mandated by CERN Council to oversee the development of a High Energy Physics Accelerator R\&D Roadmap. High Field Magnets (HFM) is one of the R\&D axes and are among the key technologies that will enable the search for new physics at the energy frontier. 

The present state of the art in HFM is based on Nb\textsubscript{3}Sn, with magnets producing fields in the range of 11 T to ~15 T. Due to the challenges associated with the brittle nature of this material, more work is required including conductor manufacturing since it is not robust enough to be considered ready at an industrial scale. Great interest has been also stirred by the recent progress achieved on HTS, not only in the fabrication of demonstrators for particle physics, but also in the successful test of magnets for other applications such as fusion and power generation. This shows that the performance of HTS magnets will exceed that of the Nb\textsubscript{3}Sn, and that the two technologies can be complementary to produce fields in the range of 20 T, and possibly higher.

The R\&D program proposed in EU ~\cite{A7-4} has two main objectives. The first is to demonstrate Nb\textsubscript{3}Sn magnet technology for large-scale deployment. This will involve pushing it to its practical limits in terms of ultimate performance (towards the 16 T target required by FCC-hh), and moving towards production scale through robust design, industrial manufacturing processes and cost reduction. The second objective is to demonstrate the suitability of HTS for accelerator magnet applications beyond the range of Nb\textsubscript{3}Sn, with a target of 20+ T. The determination of a cost-effective and practical operating field will be one of the main outcomes of the R\&D work. Sub-scale and insert coils will be used as a vehicle to demonstrate performance and operation of HTS beyond the range of Nb\textsubscript{3}Sn. The development of both a hybrid LTS/HTS accelerator magnet demonstrator, and a full HTS accelerator magnet demonstrator, is proposed. Special attention will be devoted to the possibility of operating in an intermediate temperature range of 10 K to 20 K in particular to make progress on energy efficiency of future accelerator projects.

The expected duration of the R\&D phase is 7 years. At least 5 more years will be required to develop HTS demonstrators with all the necessary accelerator features, surpassing Nb\textsubscript{3}Sn performance or working at temperatures above the liquid helium. Taking Nb\textsubscript{3}Sn as the natural reference for future accelerator magnets, HTS represents a real opportunity, provided the current trend of production and price reduction is sustainable.

Additional cross-cutting technology activities will be a key seed for innovation. The scope includes materials and composites development using advanced analytics and diagnostics, new engineering solutions for the thermal management of high-field magnets, and development of modelling tools within a unified engineering design framework. Alternative methods of detection and protection against quench (especially important for HTS) including new measurement methods and diagnostics will be explored. Finally, dedicated manufacturing and test infrastructure required for the HFM R\&D programme, including instrumentation upgrades, needs to be developed, built and operated through close coordination between the participating laboratories.
\subsection{Japan}
R\&Ds on superconducting magnets for accelerator applications has been conducted at KEK for more than 40 years. Activities involved both project-required magnets based on Nb-Ti superconductor and development of new magnet technologies with advanced superconductor such as A15 or HTS. The A15 Nb\textsubscript{3}Sn conductor development, aiming for the Future Circular Collider (FCC-hh), is now being conducted in collaboration with CERN, Tohoku Univ., Tokai Univ., NIMS, and two Japanese industrial partners. For high-field and high-radiation environment applications using HTS developments, a US-Japan collaboration was formed by KEK, Kyoto Univ., LBNL, and BNL. The quench protection as well as effects of shielding currents in the HTS conductors are also being studied.

The next mid-term goals for the superconducting magnet technologies at KEK are defined in three categories ~\cite{A7-6}: 1) high-precision 3D magnet technology based on the g-2/EDM magnet developments, 2) radiation-hard magnet technology based on the COMET magnet developments, and 3) high-field superconducting magnet technology for future colliders based on the LHC MQXA and HL-LHC D1 developments. 

Radiation-hard magnet materials and technologies are needed for future projects such as a second target station at J-PARC Material and Life Science Facility (MLF), and IR magnets for FCC-hh. To develop radiation-hard HTS magnets, R\&D is now on going within the US-Japan cooperation. The series of irradiation tests using various ReBCO conductor such as GdBCO, EuBCO, or YBCO will be compared, and most sustainable conductor will be selected. In parallel, A15 conductors are also being developed. The development of Nb\textsubscript{3}Sn conductor is now under way in collaboration with CERN, Tohoku Univ., Tokai Univ., NIMS, and two Japanese manufacturers. One of the R\&D conductors has reached 1100 A/mm$^2$ at 16 T, and the test production of 5 km level conductors with the similar specification is now in progress.
\subsection{China}
In 2012, after the discovery of Higgs boson, Chinese scientists proposed a 240 GeV Circular Electron Positron Collider (CEPC) for Higgs studies ~\cite{A7-26}. The tunnel of CEPC will also provide space for a 75-150 TeV Super Proton Proton Collider (SPPC) ~\cite{A7-28}, a candidate next-generation high-energy collider following the operation of CEPC and LHC. To reach 75-150 TeV center-of-mass energy, SPPC needs thousands of 12-24 T accelerator magnets to bend and focus the particle beams. SPPC demands advanced or new type of superconducting materials with low cost and capable of working at high fields. Since 2008, iron-based superconductors (IBS) have been discovered and attracted wide interest for both basic research and practical applications. IBS have high upper critical field beyond 100 T, strong current carrying capacity and lower anisotropy. The Institute of Electrical Engineering, Chinese Academy of Sciences (IEE-CAS) manufactured the world’s 1st 100-m long 7-filamentary Sr122 IBS tape with critical current of 1×10\textsuperscript{4} A/cm\textsuperscript{2} at 10 T successfully. This tape can now be used to fabricate IBS coils.
R\&D of high-field accelerator magnets is ongoing at IHEP in collaboration with related institutes working on fundamental sciences of superconductivity and advanced HTS superconductors. The R\&D will focus on the following key issues related to high-field superconducting magnet technology:
\begin{enumerate}
\item Exploring new methods and related mechanism for HTS materials with superior comprehensive performance for applications. Reveal key factors in current-carrying capacities by studying microstructures and vortex dynamics. Develop advanced technologies of HTS wires for high field applications with high critical current density and high mechanical strength.
\item Development of novel high-current-density HTS superconducting cables, and significant reduction of their costs. Exploration of novel structures and fabrication process of high field superconducting magnets, based on advanced superconducting materials and helium-free cooling method.
\item Search of novel stress management and quench protection methods for high-field superconducting magnets, especially for high-field insert coils with HTS conductors. Complete the prototype development with high field and 10\textsuperscript{-4} field quality, lay the foundation for the applications of advanced HTS technology in high-energy particle accelerators. The R\&D of the high-field magnet technology and related advanced superconducting materials is being funded by the Chinese Academy of Sciences (CAS) and the Ministry of Science and Technology (MOST). The total amount is about 60M\$ for 2018-2024.
\end{enumerate}
\section{HEP Accelerator Magnets Status and Future}
This section describes conceptual designs and parameters of superconducting accelerator for the electron, proton and muon colliders under consideration. The discussion starts with magnets for electron and proton colliders planned for construction in Europe and China, and concludes with magnets for muon collider under consideration in the future.
\subsection{Electron Collider Magnets} 
\subsubsection{FCC-ee magnets}
The FCC-ee main magnet requirements are quite similar to those of LEP: the arcs contain many long, low-field bending magnets interleaved with short straight sections, containing quadrupoles and auxiliary magnets ~\cite{A7-43}. The resistive magnet system resembles those used in LEP and in some other large lepton machines such as HERA electron ring and SLC. Many of its features can be retained, for example, modular cores with aluminium busbars. However, as a major innovation, for FCC-ee it is possible to exploit a dual-aperture approach. Coupling the two collider rings through twin-aperture arc magnets does not only halve the total number of main magnets, but it also saves of 50\% in electric power consumption compared with two separate sets of magnets. For operation at high energy, especially at or above the tt threshold, the SR energy sawtooth is important. To compensate for its effect, the arc dipole and quadrupole fields have to be tapered, i.e. adjusted for the local beam energy. This tapering could be achieved, for example, with individual-aperture trim coils added to the main magnets. Such trim coils for the main bending magnets could also serve as horizontal orbit correctors. 

\begin{figure}
\begin{center}
\includegraphics[width=0.60\hsize]
{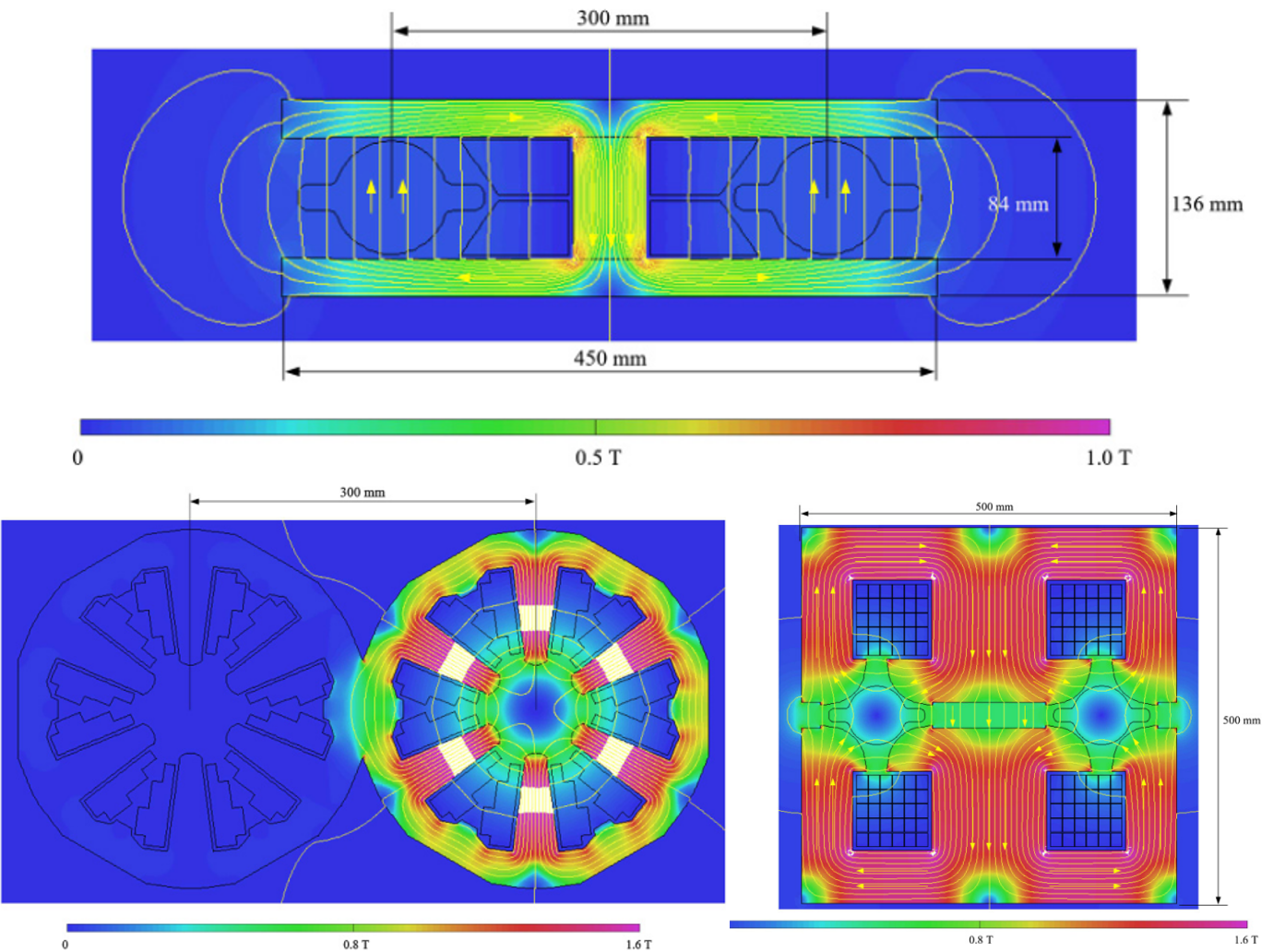}
\end{center}
\caption{FCC-ee main dipole (top), quadrupole (bottom, left) and sextupole (bottom, right) magnets.}
\label{fig:AF7-magnets-3.1}
\end{figure}

Figure \ref{fig:AF7-magnets-3.1} shows the conceptual design for the main dipole, quadrupole and sextupole magnets. For the dipoles, the design of the magnetic yoke is based on an I-type configuration, combining two back-to-back C-type layouts. In this way, the return conductor for one aperture provides the excitation current for the other. The quadrupoles cannot be considered to be low field magnets because although the beam, which is quite small with respect to the physical aperture, sees at most 100 mT (that is, 10 T/m at 10 mm), the pole tip field reaches 0.42 T. This has an impact on the Amp-turns and the power consumption. It will be even more critical for large aperture sextupoles, where the field grows quadratically from the centre. Therefore, a twin aperture layout providing significant power savings is also particularly interesting for the quadrupoles, even if they are relatively short compared to the dipoles

\begin{figure}
\begin{center}
\includegraphics[width=0.60\hsize]
{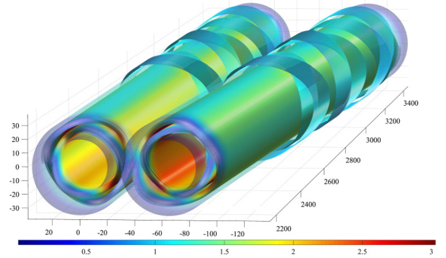}
\end{center}
\caption{Focusing quadrupoles based on the CCT technology.}
\label{fig:AF7-magnets-3.2}
\end{figure}

The Canted Cosine Theta (CCT) technology without an iron yoke has been chosen for the final focusing quadrupoles (Figure \ref{fig:AF7-magnets-3.2}). This technology provides the required field quality and has many possibilities for customisation of the field which is necessary for cross talk compensation (the tips of the FF quadrupoles closest to the IP are only 66 mm from the beams).

\subsubsection{CEPC magnets}
\begin{figure}
\begin{center}
\includegraphics[width=0.80\hsize]
{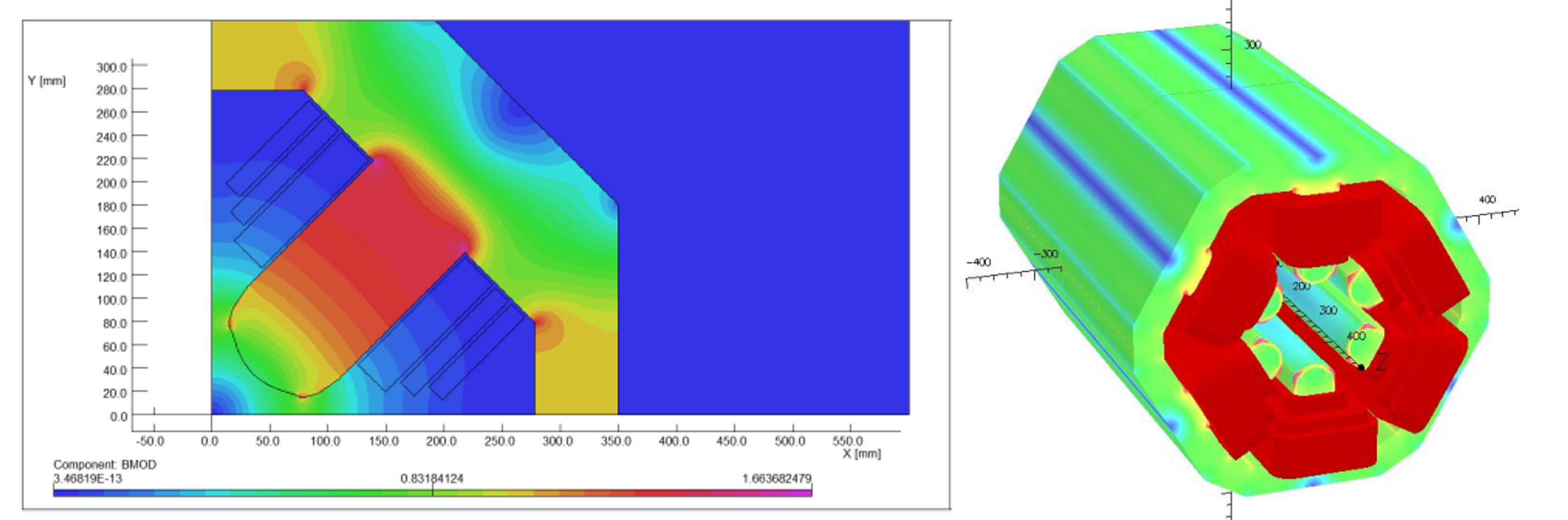}
\end{center}
\caption{CEPC quadrupole (left) and sextupole (right) magnets.}
\label{fig:AF7-magnets-3.3}
\end{figure}

The dipole field for CEPC main bending dipoles is 0.07 T. Each magnet has four C-shaped steel-concrete cores of about 4.5 m length, installed end to end in groups. The cores are composed of stacks of low carbon steel laminations, 1.5 mm thick, spaced by 6 mm gaps filled with cement mortar. The filling factor of 0.2 gives a small drop of ampere turns at the maximum field. Because of large quantity of 1,984 magnets and their long length of 18 m, the scheme of steel-concrete core not only increases the working field in the yokes of the dipole magnets but also reduces their cost significantly. The design of the quadrupoles is similar to that of quadrupole magnets in LEP. Hollow aluminium conductor is used for the coils. The iron core is made of laminated low carbon silicon steel sheet with a thickness of 1.5 mm. The pole has parallel sides so that simple racetrack shaped coils can be used. The magnet will be assembled from four identical quadrants and can also be split into two halves for the installation of the vacuum chamber. There are two types of sextupole magnets in the Main Ring, with the same aperture and cross section but of different lengths. The cores of the sextupole magnets are made from 1.5 mm thick low carbon steel laminations. To reduce cost, the coils will be wound from hollow aluminium conductors. Figure \ref{fig:AF7-magnets-3.3} shows a cross section of the quadrupole magnet and the 3D view of the sextupole ~\cite{A7-30}. 

\subsection{High Field Proton Collider Magnets} 
High field magnets are critical to energy-frontier circular colliders. For the main ring dipole magnets of a hadron collider, the linear relationship between beam energy, ring radius, and magnetic field strength provide the basis for facility optimization for physics reach and for cost. A reasonable estimate for the beam energy is E = 0.3$\cdot$B$\cdot$R, with energy E in TeV, field B in T, and radius R in km. The maximum affordable dipole field, together with the dimension of the tunnel, set the maximum achievable energy, so the focus is set on the development of cost-effective high-field accelerator dipoles. In the insertion and straight section regions, where the experiments are housed, special magnets are needed for the strong focusing of the beam in the interaction point or for the crossing in the beam. Nb\textsubscript{3}Sn is today the natural reference for future collider magnets, but HTS has the potential to operate at higher fields (20 T target) and/or intermediate temperature range of 10 to 20 K. 
\subsubsection{FCC-hh magnets}
\begin{figure}
\begin{center}
\includegraphics[width=1.0\hsize]
{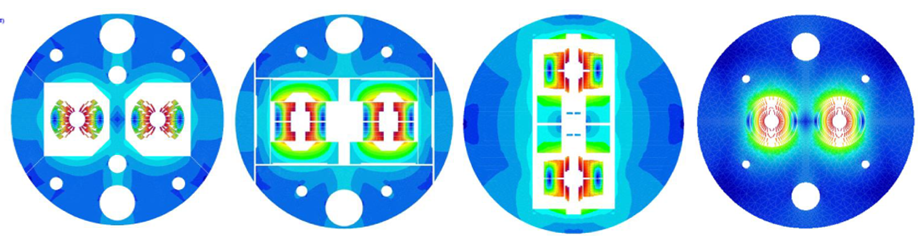}
\end{center}
\caption{Cross section of the four main dipole design options explored within the EuroCirCol program. From left to right, cos-theta (INFN), bock (CEA), common coil (CIEMAT), and CCT (PSI) designs.}
\label{fig:AF7-magnets-3.4}
\end{figure}

A Future Circular Collider (FCC) require bending magnets operating at up to 16 T~\cite{A7-29}. This is about twice the magnetic field amplitude produced by the Nb-Ti LHC magnets, and about 5 T higher than the pole tip field of the Nb\textsubscript{3}Sn magnets being developed for the High Luminosity LHC (HL-LHC). The design of the magnets was explored within the WP5 EuroCirCol Program. A specific feature of the program is that different design options are considered with the same specification, to provide a 16 T dipole field in a 50 mm aperture. The target critical current density is 1500 A/mm2 at 4.2 K and 16 T, corresponding to 2300 A/mm2 at 1.9 K and 16 T. The margin on the load line was at least 14\% for all designs and the maximum allowable stress in the conductor was 150 MPa at room temperature and 200 MPa at 1.9 K. A considerable effort has indeed been devoted to finalize optimum common design parameters and assumptions, as well as setting up and validating the design and analysis tools, in particular the ones for quench protection. Since the beginning of the activity in July 2015, the program went through several major evolutions. In 2016 a redefinition of the conductor and load line margin parameters allowed the design of electromagnetically efficient coil cross sections, new features implemented in the common coil design ~\cite{A7-35} allowed to partially fill the gap of amount of conductor needed with respect to the cosine-theta ~\cite{A7-33} and the block-coil ~\cite{A7-34} configurations, and finally a new concept of canted-cos-theta using large cables was introduced ~\cite{A7-36}. The electromagnetic cross section of each of these options is shown in  Figure \ref{fig:AF7-magnets-3.4}. Their salient features, with respect to the baseline cosine-theta, are shown in Table \ref{tab:AF7-magnets-table-3.1}.

\begin{table}
\begin{center}
\caption{\label{tab:AF7-magnets-table-3.1} Parameters of design options for the 16 T arc dipole.}
\begin{tabular}{ |c|c|c|c|c| } 
 \hline
Parameter & Cos-theta & Block-coil & CCT & Common-coil\\ 
\hline
Peak field on conductor (T) & 16.40 & 16.73 & 16.35 & 16.57 \\ 
Operating current (A) & 11441 & 10176 & 18135 & 15880 \\ 
Inductance @ 16 T (mH/m) & 38 & 48 & 18 & 26 \\
Outer yoke diameter (mm) & 660 & 616 & 750 & 650 \\
Mass of conductor (kg/m) & 115 & 120 & 148 & 145 \\
 \hline
\end{tabular}
\end{center}
\end{table}

\begin{figure}
\begin{center}
\includegraphics[width=0.6\hsize]
{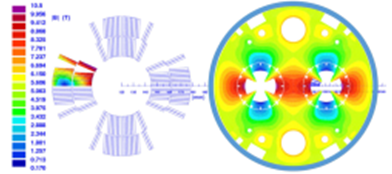}
\end{center}
\caption{Left: coil cross-section for the FCC main quadrupoles (MQ) with magnetic field map. Right: cold-mass cross-section.}
\label{fig:AF7-magnets-3.5}
\end{figure}

The FCC main quadrupoles (MQ) are twin-aperture magnets based on a cos-theta coil configuration assembled in a 20 mm thick helium II vessel ~\cite{A7-37}. Like the main dipole magnet, the inter-beam distance is 250 mm and the physical aperture is 50 mm in diameter. Each aperture is mechanically independent from the other due to the use of a collar-and-key mechanical assembly. Each double pancake is made of 18 turns of Nb\textsubscript{3}Sn Rutherford cable with a 0.4 degree keystone angle. The electromagnetic cross-section is shown in  Figure \ref{fig:AF7-magnets-3.5}.
For the FCC-hh, the FODO cell length in the arc is 213 m, roughly double the length of the LHC FODO cell. The FCC-hh has 8 long arcs, each with 36.5 FODO cells, and 4 short arcs, each with 15 FODO cells. Each FODO cell has 12 dipoles and 2 Short Straight Sections (SSS). As in the LHC, each SSS contains one quadrupole MQ, and sextupole (MS) and dipole corrector magnets (MC). Depending on the SSS location in the arc, there may be in addition octupole corrector magnets (MO), tuning quadrupoles (MQT) or skew quadrupoles (MQS). The magnet types and their main parameters are listed in Table \ref{tab:AF7-magnets-table-3.2}. All these magnets, except the DIS quadrupole (MQDA), use Nb–Ti technology ~\cite{A7-38}, whereas the MQDA relies on Nb\textsubscript{3}Sn.
\begin{table}
\begin{center}
\caption{\label{tab:AF7-magnets-table-3.2} Other magnets in the FCC-hh arc.}
\begin{tabular}{ |c|c|c|c|c| } 
 \hline
Magnet type & Technology & Number & Strength & Length\\ 
\hline
Lattice sextupoles (MS) & Nb-Ti & 696 & 7000 T/m$^3$ & 1.2 m \\ 
Lattice octupoles (MO) & Nb-Ti & 480 & 200000 T/m$^3$ & 0.5 m \\ 
Dipole correctors (MCB) & Nb-Ti & 792 & 4 T & 1.2 m \\
Trim quadrupoles (MQT) & Nb-Ti & 120 & 220 T/m & 0.5 m \\
Skew quadrupoles (MQS) & Nb-Ti & 96 & 220 T/m & 0.5 m \\
Sextupole spool correctors (MCS) & Nb-Ti & 2 x 4668 & 3000 T/m$^2$ & 0.11 m \\
DIS quadrupole (MQDA) & Nb$_{3}Sn$ & 48 & 360 T/m  & 9.7 m \\
DIS trim quadrupole (MQTL) & Nb-Ti & 48 & 220 T/m & 2.0 m \\
 \hline
\end{tabular}
\end{center}
\end{table}

The low-beta triplets are composed of quadrupole magnets and corrector magnets. There are two types of low-beta triplets for installation in the high- and low luminosity interaction regions, respectively. The magnet types and their main parameters are listed in Table \ref{tab:AF7-magnets-table-3.3}. 
The magnets around the collision points will be exposed to high radiation levels which may adversely affect their performance. It is assumed that the conductor performance can be maintained until a displacement-per-atom (DPA) value of $2 \cdot 10 ^{-3}$ and that the magnet insulation can withstand an accumulated radiation dose of 30 MGy. These values may be exceeded over the machine lifetime, going up to 40–50 MGy assuming the use of the baseline 35 mm thick tungsten shield. 
\begin{table}
\begin{center}
\caption{\label{tab:AF7-magnets-table-3.3} Low-beta triplet magnets.}
\begin{tabular}{ |c|c|c|c|c|c| }
 \hline
Magnet type & Technology & Number/IP & Strength & Length & Aperture\\ 
\hline
Q1 high lumi & Nb$_{3}$Sn & 4 & 130 T/m & 14.3 m & 164 mm \\ 
Q2 high lumi & Nb$_{3}$Sn & 8 & 105 T/m & 12.5 m & 210 mm \\ 
Q3 high lumi & Nb$_{3}$Sn & 4 & 105 T/m & 14.3 m & 210 mm \\
Q1 low lumi & Nb-Ti & 4 & 270 T/m & 10.0 m & 64 mm \\
Q2 low lumi & Nb-Ti & 8 & 270 T/m & 15.0 m & 64 mm \\
Q3 low lumi  & Nb-Ti & 4 & 270 T/m & 10.0 m & 64 mm \\
 \hline
\end{tabular}
\end{center}
\end{table}

\subsubsection{SPPC magnets}
The SPPC design study is considering accelerator dipole magnets with nominal magnetic fields in the range of 12-24 T, an aperture of 40-50 mm and the field uniformity on the level of $10^{-4}$ in ~65\% of their aperture. The magnets have two apertures with opposite field direction inside the common iron yoke to minimize the magnet transverse size and reduce its cost. The aperture separation in the main dipoles is presently estimated on the level of 200-300 mm. This parameter in the final magnet design will be optimized to achieve the acceptable crosstalk between two apertures and minimize the overall magnet cross-section. The outer diameter of the arc dipole and quadrupole cold masses should be below by 900 mm in order to be installed inside vacuum vessels with an outer diameter of 1.5 m. The total magnetic length of the main dipoles is ~65.4 km at the total collider ring circumference of 100 km. For the dipole length of about 15 m, approximately 4360 dipole magnets will be needed ~\cite{A7-39}. 

\begin{figure}
\begin{center}
\includegraphics[width=0.6\hsize]
{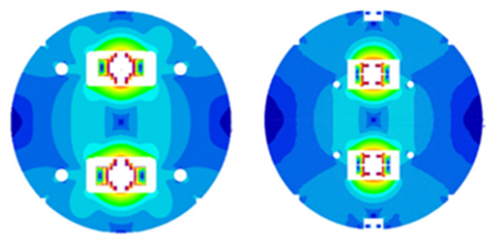}
\end{center}
\caption{Cross-sections of the two design options for the 12 T SPPC dipole based on IBS.}
\label{fig:AF7-magnets-3.6}
\end{figure}

SPPC magnets with operation fields up to 15-16 T will need advanced Nb\textsubscript{3}Sn superconductors. To increase the nominal operation fields up to 24 T, High Temperature Superconducting (HTS) materials with acceptable cost are needed. Special attention at the present time is being paid to Iron Based Superconductors (IBS) discovered in 2008 which show potential for improved performance at lower cost. Conceptual design studies of twin-aperture 12 T dipole magnets for the SPPC based on the IBS technology are being performed. Cross-sections of the two design options under consideration for the 12 T SPPC dipole based on the IBS conductor are shown in Figure \ref{fig:AF7-magnets-3.6}. The coil aperture in both cases is 45 mm and the nominal field in the two magnet apertures is 12 T with the geometrical field quality on the level of $10^{-4}$. The coil layout uses the common-coil configuration due to its simple structure compatible with the brittle conductor and simplicity for fabrication.
\subsection{Muon Collider Magnets}
\subsubsection{Muon collider magnet system}
\begin{figure}
\begin{center}
\includegraphics[width=0.9\hsize]
{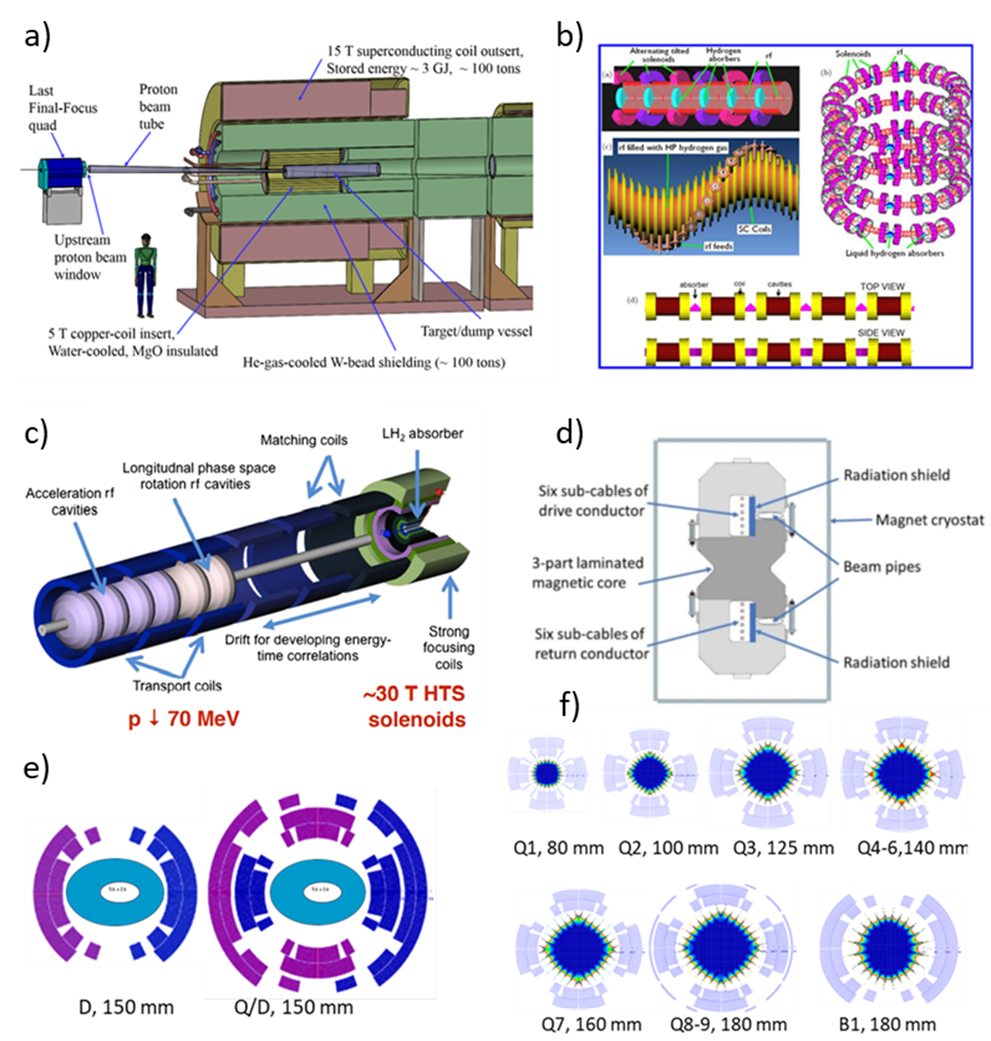}
\end{center}
\caption{MC magnets: a) production solenoid; b) 6D muon cooling solenoids: c) final muon cooling system; d) fust-cycling muon acceleration magnet; e) collider ring magnets; f) IR magnets.}
\label{fig:AF7-magnets-3.7}
\end{figure}

Muon colliders are a promising path to reach very high center of mass energies. Muon Collider (MC) magnet needs and challenges are discussed in ~\cite{A7-5}. The pictures of main MC magnets are shown in  Figure \ref{fig:AF7-magnets-3.7}. The results achieved in the framework of the U.S. Muon Accelerator Program (MAP) ~\cite{A7-40}, the International Muon Collider Collaboration (IMCC), and the Snowmass Muon Collider Forum ~\cite{A7-41} provide a broad view of the percived challenges and technology options.
\subsubsection{Muon production}
Muons are produced by the decay of pions generated by interaction of intense proton beams with a target placed inside a high-field production solenoid. This solenoid captures the pions and guide them into a decay channel to produce muons. The production solenoid produces a high magnetic field of 20 T in a 150 mm aperture. A present baseline option is based on the 12-15 T, 2.4 m bore superconducting (SC) outsert solenoid and the 5-10 T, 150 mm bore normal conducting (NC) insert. As an alternative, a large bore LTS and HTS hybrid solenoid is also considered, although high resistance to radiation and heat load of this design would need to be demonstrated. The key advantages of an HTS-based solution would be the reduction of the outer diameter, the large temperature margin of HTS, the possibility to operate at higher temperature than liquid helium, and the reduction of the operation costs. The main challenges of the production solenoid include high operation field, strong electromagnetic forces, large stored energy, high-radiation environment. While an LTS+NC hybrid solenoid, originally planned in MAP, can be built based on the extrapolation of known technology, an HTS+NC or LTS+HTS hybrid versions would require a comprehensive R\&D. The requirements and technological challenges of the production solenoid broadly overlap with those of magnets for high-magnetic field science, as well as solenoids for Tokamaks. The technology for a LTS+HTS option has overlap with that of ultra-high-field NMR magnets, and such development would profit from synergy with the development of HTS coils for High Energy Physics (e.g. HEP dipoles) and light sources (e.g. super bends).
\subsubsection{Muon cooling}
The cooling of muon beams occurs in a channel formed by solenoids and RF cavities. Since the final emittance of the muon beam is inversely proportional to the magnetic field of the final cooling solenoids, the goal is to produce a field of 40-60 T in a 50 mm aperture. The preferred design approach in this case is an LTS+HTS hybrid magnet. The main challenges of this ultra-high field solenoid are the large electromagnetic forces and stresses, quench protection as well as the mechanical, thermal and powering integration of the LTS and HTS coils. These solenoids go beyond available technologies and, thus, will require considerable R\&D and demonstration efforts. The challenges for the final cooling solenoids are similar to magnets for high-field magnet user facilities based on all-superconducting solenoids, and ultra-high field NMR solenoids. The R\&D on the final cooling solenoids is also synergetic with the HTS accelerator magnet R\&D for High Energy Physics (HEP) and light sources.
\subsubsection{Acceleration}
After cooling, muon beams have to be rapidly accelerated to relativistic momentum to extend their laboratory lifetime. After an initial acceleration stage in a Linear Accelerator (LINAC) and Recirculating Linear Accelerators (RLA) a sequence of Fixed Field Alternating Gradients (FFAG) Rapid-Cycled Synchrotrons (RCS) and Hybrid Cycled Synchrotrons (HCS) is used. The preferred option at the present time is RCS and HCS based on either NC fast ramping magnets (RCS) or a combination of NC fast ramping and static SC magnets (HCS). Both synchrotrons require fast ramping magnets. In the front end HCS, the fast ramped magnets for the muon acceleration need a magnetic field sweep of ~4 T (±2 T) within 0.4 ms, which corresponds to a field ramp rate of 10 kT/s, in a rectangular bore of 80 mm by 40 mm. A resistive approach is preferred for these specifications. The far end HCS require 4 T change over ~10 ms, or a field ramp rate of ~400 T/s, which makes HTS promising. A larger swing of field would be beneficial, which may be also the key advantage of HTS. Besides the engineering challenges of such magnets, the stored energy of an accelerator ring of the appropriate size is large, of the order of several tens of MJ, and operation at high-pulse rate requires managing of peak power in the range of tens of GW. Energy storage based on capacitor banks seems to be the only viable solution, or alternatives such as SMES and flux-pumps with high Q-factor to improve the energy efficiency. The work required on magnets and powering for the accelerator stage has clear synergies with the design of RCS for nuclear physics machines, as well as accelerator driven transmutation and fission systems.

A practical opportunity to build fast-cycling dipole magnets using second generation (G2) HTS is discussed in ~\cite{A7-20}. Besides being superconducting at relatively high temperatures, rare-earth HTS (REBCO) tapes have shown very small AC losses compared to low-temperature Nb-Ti cables. Recent tests of the 0.5 m long twin-aperture HTS-based dipole demonstrated record-high field ramp rate of ~300 T/s at 10 Hz repetition rate and 0.5 T field amplitude. No temperature rise in 6 K cooling He was observed within the ~0.003 K error which is setting the upper limit on the cryogenic power loss in the magnet conductor coil below 0.2 W/m. Based on this result a possible upgrade of this magnet design to 2 T in the 10 mm beam gap with the field ramp rate up to 1000 T/s is discussed.
\subsubsection{Collision}
The final stage of the muon collider complex is a storage ring (SR) and interaction region (IR). The SR needs to have the smallest possible size to maximize the number of muon collisions within their limited lifetime. Radiation shielding, from the collisions, and neutrino flux mitigation, due to muon decay, need to be addressed. As a result, the SR and IR magnets have to produce high fields in large apertures for shielding and, in the case of the IR magnets, to achieve high luminosity. For a 10 to 14 TeV muon collider, SR dipole magnets are required to generate a steady-state magnetic field up to 16 T in a 150 mm aperture. To reduce straight sections, and mitigate radiation dose, the collider magnets are presently assumed to have combined functions (D+Q and D+S). SR magnets need to sustain a heat load of 500 W/m from muon decay and synchrotron radiation. In a 3 TeV c.o.m. MC, IR quadrupoles with steady-state gradient of 250 T/m in an 80 mm aperture and 125 T/m in 180 mm aperture (coil peak field up to 12 T), are foreseen for the final beam focus. Adding appropriate operation margins, these parameters are marginally within reach of Nb\textsubscript{3}Sn. At higher energies the required dipole field and field gradient will need to be increased to the level of 20 T which would require using HTS superconductors. The main challenges of SR and IR magnets are the high field and large aperture required, as well as the radiation dose and heat load. These challenges are broadly shared with the development of high-field magnets for future collider, and in particular the need to manage stress in compact windings with high engineering current density. 

The results of 3 TeV c.o.m. MC studies and a design concept of the 6 TeV c.o.m. MC optics, the SC magnets, and a preliminary analysis of the protection system to reduce radiation loads on the magnets and particle backgrounds in the MC detector are discussed in ~\cite{A7-9}. The SC magnets and detector protection considerations impose strict limitations on the design of the collider optics, magnets and Machine Detector Interface (MDI). SC magnets based on traditional cos-theta coil geometry and Nb\textsubscript{3}Sn superconductor were used to provide realistic field maps for the analysis and optimization of the arc lattice and IR design, as well as for studies of beam dynamics and magnet protection against radiation. An important issue to address is the stress management in the coil to avoid substantial degradation or even damage of the brittle SC coils. Stress management concepts for shell-type coils are being developed for high-field accelerator magnets based on LTS (Nb\textsubscript{3}Sn) and HTS (Bi2212 and REBCO) cables by the US MDP (see section 2.1.2).

A low-energy medium-luminosity MC as a possible Higgs Factory (HF) has been also studied and is discussed in ~\cite{A7-10}. Electrons from muon decays will deposit more than 300 kW in the HF SR, which imposes significant challenges to SC magnets used in the HF SR and IR. The distribution of heat deposition in the HF SR lattice elements requires large-aperture magnets to accommodate thick high-Z absorbers to protect the SC coils. Magnet conceptual designs which provide the required high operating gradients and dipole fields in large apertures to accommodate the large size of muon beams, as well as a cooling system to intercept the large heat deposition from the showers induced by decay electrons, were developed. In the IR, the coil aperture of the IR magnets varies from 320 mm in Q1 and to 500 mm in Q2–Q4 and B1. Outside of the IR, the coil aperture varies between 270 mm and 160 mm. The maximum design field in coils in the HF SR and IR magnets is on the level 16-18 T to provide the necessary operation margin. High field level and large magnet apertures require using the Nb\textsubscript{3}Sn technology and stress management approach. A sophisticated radiation protection system was designed for the SR and IR to bring the peak power density in the SC coils below the quench limit and reduce the dynamic heat deposition in the cold mass by a factor of 100. The system consists of tight tungsten masks in the magnet interconnect regions and elliptical tungsten liners in the magnet aperture optimized individually for each magnet.

\section{Detector Magnets Progress and Future Requirements}
Superconducting detector magnets are a key component to analyze the momentum and polarity of charged particles in particle physics experiments ~\cite{A7-19}. A large warm bore is required. 

Over the past four decades, the design and fabrication of detector magnets has made significant progress thanks to the introduction and demonstration of specific technologies such as: high current, high strength, aluminum-stabilized superconducting cables; coil winding performed directly in the outer support cylinder to support large hoop stress; improved transparency using indirect cooling to eliminate the liquid helium vessel, and no mechanical structure in the winding; pure aluminum strips as temperature equalizer and for fast quench propagation. 

\begin{figure}
\begin{center}
\includegraphics[width=0.9\hsize]
{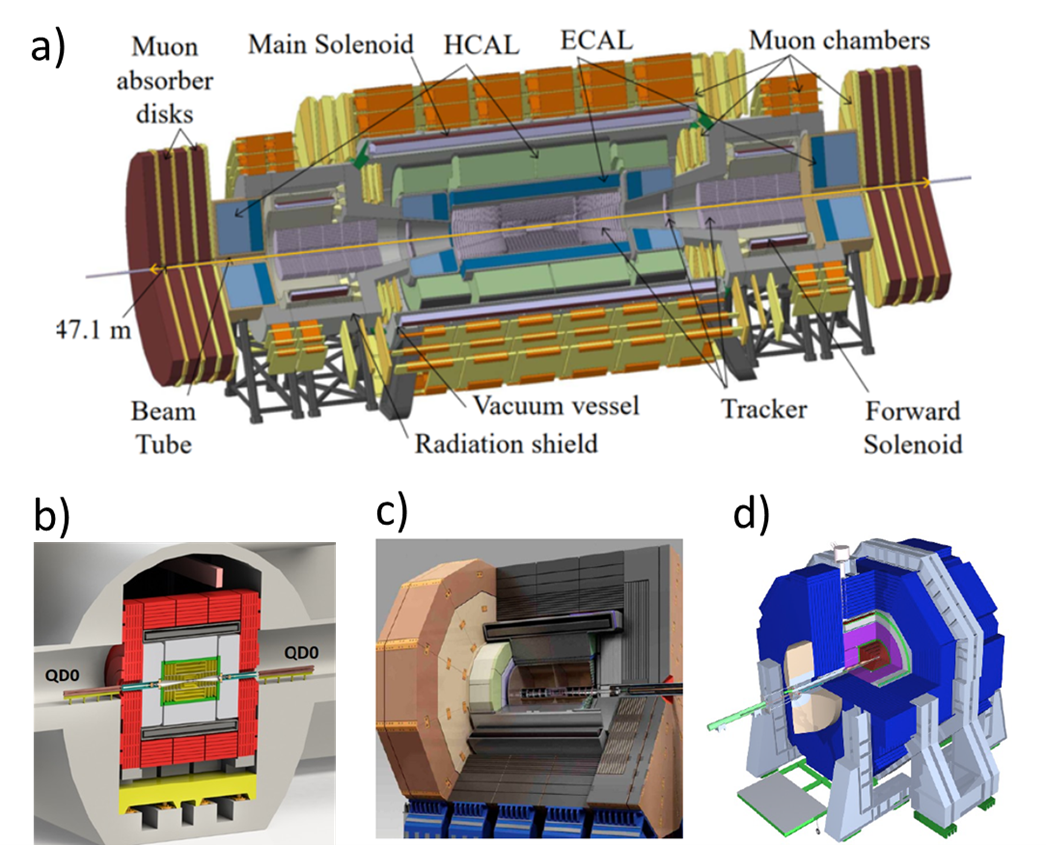}
\end{center}
\caption{Proposed detector base-line layouts: a) FCC-hh detector; b) CLIC; c) ILD; and d) SiD.}
\label{fig:AF7-magnets-4.1}
\end{figure}

Detector solenoid design studies are in progress for future big projects in Japan and Europe, that is, ILC, FCC and CLIC ~\cite{A7-19}. The proposed detector layouts are shown in  Figure \ref{fig:AF7-magnets-4.1}, design parameters for each solenoid are summarized in Table \ref{tab:AF7-magnets-table-4.1}. Detector magnets for future colliders will be based on these technologies but further improvements will be required to cope with larger size, radiation loads and physics performance requirements. At the same time there is a risk that expertise and industrial capabilities are gradually lost due to the long time between projects. The industrial technology of Al-stabilized superconductor production is of particular concern. In other cases, precise control of the magnetic field distribution requires high precision simulation, and tight control of fabrication tolerances. If the calorimeter is placed outside the magnet, such as ATLAS, high transparency is also required for charged particles passing through with a minimum energy. Otherwise, particle detectors need to be installed inside the magnet bore (except for muon detectors) resulting in larger bore, higher field and longer coil length. 
\begin{table}
\begin{center}
\caption{\label{tab:AF7-magnets-table-4.1} The proposed design paramters for the detector solenoids for future projects.}
\begin{tabular}{ |c|c|c|c|c|c|c| }
 \hline
Project & Magnet & B$_{c}$ (T) & R$_{in}$ (m) & Length (m) & E/M (kJ/kg) & Stored Energy (GJ)\\ 
\hline
FCC-ee & IDEA & 2 & 2.24 & 5.8 & 14 & 0.17 \\ 
FCC-ee & CLD & 2 & 4.02 & 7.2 & 12 & 0.6 \\ 
FCC-hh & & 4 & 5 & 20 & 11.9 & 13.8 \\
CLIC &   & 4 & 3.65 & 7.8 & 13 & 2.3 \\
ILC & ILD & 4 & 3.6 & 7.35 & 13 & 2.3 \\
ILC  & SLD & 5 & 2.5 & 5 & 12 & 1.4 \\
 \hline
\end{tabular}
\end{center}
\end{table}
\subsection{FCC-ee}
Three detector designs have been proposed for FCC-ee ~\cite{A7-42}: the Innovative Detector for Electron-positron Accelerators (IDEA, ~\cite{A7-43}), the CLIC-Like Detector (CLD, ~\cite{A7-44}) and a design comparable to the IDEA detector that remains to be named. Each of these three designs includes a 2 T superconducting solenoid. The CLD solenoid is positioned outside of the calorimeter while the two other solenoids are situated inside the calorimeter barrels. Since the IDEA solenoid is inside the calorimeter there are strict requirements on the particle transparency. The concept of the solenoid of the IDEA is similar to the ATLAS Central Solenoid (CS) ~\cite{A7-45} whereas the CLD solenoid is similar to the CMS solenoid ~\cite{A7-46}. However, the free-bore diameter of the IDEA solenoid is almost two times bigger than the free bore of the ATLAS CS. This also means that the IDEA magnet has around four times the stored magnetic energy of the ATLAS CS at 170 MJ. The free-bore diameter of CMS is 6 m with a stored energy of 2.6 GJ. The CLD design has a larger free bore of 7.2 m and its stored energy is 600 MJ.  

The challenges of the IDEA and CLD designs are well illustrated by their energy density. The energy density in IDEA solenoid is 14 kJ/kg, which is twice as high in ATLAS, whereas the CLD magnet has an energy density of 12 kJ/kg which is close to 11 kJ/kg in CMS. The large free bores in combination with the high stored energies translate to strong requirements on quench protection design, mechanical support and strength of the materials used. Preliminary protection studies are using energy extraction in combination with quench heaters and aluminum strips for faster propagation. For both detectors a safe peak temperature of about 60 K can be maintained.   
\subsection{FCC-hh}
The FCC-hh ~\cite{A7-50} detector magnet features three 4 T superconducting solenoids~\cite{A7-49}. The central solenoid has 10 m diameter bore and a length of 20 m, whereas the forward solenoids have a cold mass length of 3.4 m and a free bore diameter of ~5 m. The basic detector layout is similar to CMS, featuring a tracker, electromagnetic calorimeter (E-CAL), and a hadron calorimeter (H-CAL) in the bore of the magnet, and muon chambers on the outside of the magnet. The muon chambers utilize the magnetic return flux of the main solenoid for the purpose of muon tagging. However, unlike CMS, the FCC-hh detector does not feature iron yokes.

The bore field of 4 T and the energy density of 11.85 kJ/kg are a bit higher than CMS, while the total stored energy of 13.8 GJ is more than five times higher. The peak Von Mises stresses in the main and the forward solenoids are at 100 MPa under nominal conditions which shows that, similar to CMS and ATLAS Central Solenoid, a reinforced conductor is needed to handle the Lorenz forces. The conductors are made of Nb-Ti/Cu Rutherford cables surrounded by nickel-doped aluminum stabilizer. The cables feature 40 strands with a diameter of 1.5 mm, a Cu/non-Cu ratio of 1:1, and a current sharing temperature of 6.5 K. The operating current is 30 kA.

The solenoids are powered in series by a single power supply with slow dump circuit which is sufficient to charge and discharge the solenoids. In case of a quench, the different operating current densities in the main and forward solenoids necessitate a decoupling of these two different magnet types, and therefore each type features their own fast discharge dump system comprising diodes and resistors. In parallel, protection heaters are used to initiate quenches in different areas of the coils, thus avoiding strong temperature gradients even under fault conditions where the fast dump units fail to discharge the magnets. The calculated peak hot-spot temperature is well below 100 K under nominal conditions.
\subsection{CLIC}
The CLICdet design ~\cite{A7-44} is based on a 4 T solenoid and is designed to operate at the three stages of the CLIC accelerator phases with c.o.m. energies of 380 GeV, 1.5 TeV, and 3 TeV. It takes advantage of the technical approaches developed for the CMS solenoid ~\cite{A7-51} and the Atlas Central Solenoid (CS) ~\cite{A7-52}. The conductor comprises a Rutherford cable composed of 32 Nb-Ti strands that is stabilized with an aluminum sheath. Both ATLAS CS and CMS approaches for a reinforced conductor are being considered at this stage. The Rutherford cable will be co-extruded with the high Residual Resistivity Ratio (RRR) aluminum stabilizer. The end coils are used both to limit the magnetic field in the machine detector interface region, in particular on the QD0 final focusing quadrupoles located just in front of the end caps, and to limit the stray field outside the detector to 16 mT at a radial distance of 15 m in the service cavern The use of these ring coils also contributes to limit the amount of iron in the yoke. It was proposed to build these end coils with normal conducting windings operated at room temperature and water cooled, but there are opportunities for more sustainable solutions by using high temperature superconductor coils in order to limit the heat losses due to the dissipated power.

Several technology R\&D and pre-industrialization programs will be needed before launching the manufacturing of such a magnet. Power leads and superconducting busbars are typical applications that can be developed using high temperature superconductors. Other developments can also be performed for powering DC converters and dumping circuit. Dedicated studies applied to the detector magnet applications will be needed.

\subsection{ILC}
Two detectors are proposed for ILC: ILD and SID. The design parameters of the magnet for the International Large Detector (ILD) feature a central field of up to 4 T, with useful diameter 6.88 m over a length of 7.35 m. Its conceptual design has been done by CEA, DESY and CERN. Many technical solutions successfully used for CMS are proposed for the ILD magnet. Similarly to CMS, a 4-layer coil is retained, with a nominal current of 20 kA, so the conductor for the ILD magnet has larger cross-section of 74.3×22.8 mm$^2$. It consists of a superconducting Rutherford cable, clad in a stabilizer and mechanically reinforced. Two solutions are considered for the reinforcement. The first option is a micro-alloyed material such as the ATLAS central solenoid, which acts both as a stabilizer and a mechanical reinforcement. A R\&D program on the Al-0.1wt\%Ni stabilizer has been launched at CERN and is underway to demonstrate the feasibility of producing a large conductor cross section with this material. The second option is a CMS-type conductor with two aluminum alloy profiles welded by electron beam to the central conductor stabilized with high purity aluminum. 

A conceptual design study for a 5 T superconducting solenoid for the ILC Silicon Detector (SiD) has been done by FNAL. The solenoid has a clear bore diameter of 5 m and a length of 5 m, where 5 T magnetic field is produced for inner detectors. The coil utilizes the CMS conductor as the starting point. Major R\&D subject for SiD magnet is its conductor required to sustain large EMF in the coil. A more advanced and likely cheaper conductor was proposed based on high-purity aluminum Al-0.1\%Ni alloys, which were used in the ATLAS central solenoid. Other conductor stabilizer possibilities are also under consideration and study. These include TiB2 grain refinement aluminum matrix composites, and cold working via the equal area angle extrusion process.

\section{Magnet technology R\&D for High Energy Physics}
\subsection{Superconducting composites}

\begin{figure}
\begin{center}
\includegraphics[width=0.8\hsize]
{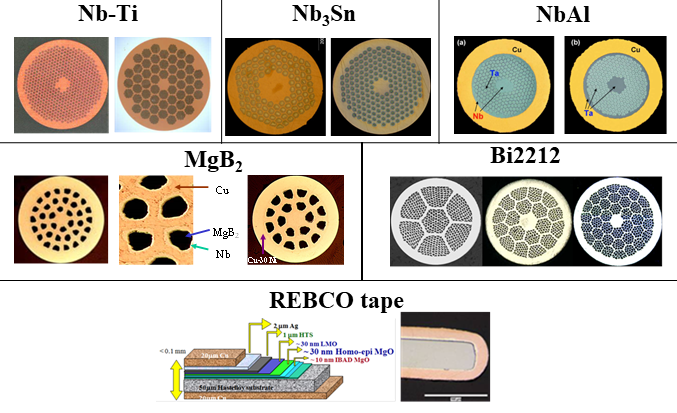}
\end{center}
\caption{Composite SC wire and tapes.}
\label{fig:AF7-magnets-5.1}
\end{figure}

There is a large variety of technical superconductors with critical parameters, such as critical temperature T\textsubscript{c}, upper critical field B\textsubscript{c2} and critical current density J\textsubscript{c}, sufficient for using in large magnet systems. Important features of practical materials for superconducting magnets include performance (appropriate combination of critical parameters) and its reproducibility in long lengths, commercial production and affordable cost. The superconducting materials described below belong to the class of practical superconductors being used in various accelerator and detector magnets. Table \ref{tab:AF7-magnets-table-5.1} summarizes the critical performance and production parameters of the superconductors available for procurement in industry in 1 km length that also meet needs for accelerator magnets.  Cross-sections of the practical wires and tapes are shown in  Figure \ref{fig:AF7-magnets-5.1}. Magnet conductors are discussed in~\cite{A7-8}.
\begin{table}
\begin{center}
\caption{\label{tab:AF7-magnets-table-5.1} Critical performance paramters and productions parameters of the conductors.}
\begin{tabular}{ |p{1.5cm}|p{1.5cm}|p{1.5cm}|p{1.5cm}|p{1.5cm}|p{2cm}|p{1.5cm}|p{1.5cm}| }
 \hline
SC material & T$_c$, K & B$_{c2}$ (4.2 K) & Mechanical property & Billet or batch mass, kg & Annual production scale, tons & Relative cost & Final cost /material cost\\ 
\hline
Nb-Ti & 9.2 & 11 & ductile & 200-400 & $>$100 & 1 & 3 \\ 
Nb$_{3}$Sn & 18 & 26 & brittle & 45 & 1-10/1* & 5/8* & 5-7 \\ 
Nb$_{3}$Al & 18 & 26 & brittle & 45 & 1 & 8 & \\
MgB$_{2}$ & 39 & 20 & brittle & 20 & $<$1 & 2 & \\
Bi2212 & 85 & 100 & brittle & 20 & $<$1 & 20-50 & $\approx$ 10 \\
REBCO  & 92 & $>$120 & brittle & 10 & $<$1** & 20-50 & $>>$10 \\
 \hline
\end{tabular}
\end{center}
{\raggedright *RRP/PIT
** few tons planned for fusion \par}
\end{table}


\subsubsection{LTS composite wires}
\underline{Nb-Ti composite wires.} Since its discovery in the 1960s, Nb-Ti is the most successful practical superconductor. A composite Nb-Ti wire is manufactured by a co-drawing process with intermediate heat treatments to achieve an optimal pinning structure. Diameters of practical wires are in the range of 0.1 to 3 mm and piece lengths are of a few km. The typical RRR value of the Cu matrix is in the range of 50–200. A thin Nb barrier separates the Nb-Ti alloy from the Cu to avoid the formation of brittle CuTi intermetallic composite during the high-temperature extrusion and the annealing heat treatments. The number of filaments in a wire may reach ~105 and the filament diameter is 5–50 µm (the practical low limit is ~1–2 µm). The number of filaments and, thus, the filament size are determined mainly by stability criteria and by hysteresis magnetization or AC loss values.  Figure \ref{fig:AF7-magnets-5.1} shows a typical cross-sections of Nb-Ti superconducting composite wires. The single stack design is limited to ~104 filaments. Using the double stack design allows achieving larger-number and smaller-size filaments.

Artificial Pinning Centers (APC) based on thin normal metal rods incorporated in the Nb-Ti filaments were proposed in the 1970s to improve and control flux pinning. This method has resulted in enhanced performance at lower fields with respect to conventional Nb-Ti, as needed in some magnet applications. It was found also that ternary Nb-Ti-Ta alloys have the potential to provide a small gain in upper critical field (up to 1.25 T at 1.8 K) while preserving the traditional wire manufacturing process. Low-loss Nb-Ti composite wires with ultrafine filaments (~0.1 µm) and extra resistive Cu-Ni matrix are also produced for AC applications.

\underline{Nb\textsubscript{3}Sn composite wires.} Nb\textsubscript{3}Sn is an intermetallic composite, the second practical superconductor most widely used in superconducting magnets. Its superconducting properties were discovered in the 1960s. Nb\textsubscript{3}Sn composite wires are currently produced using three main methods: bronze, internal tin (IT), and powder-in-tube (PIT) methods. The last two are most interesting for accelerator magnets due to high Jc.

The Nb\textsubscript{3}Sn wires produced by IT Restacked-Rod-Process (RRP) use an architecture, wherein rods of Nb, Nb-Ta alloy, or Nb-Ti alloy are sheathed in Cu and extruded, drawn, stacked in an annulus around a central core, wrapped with a diffusion barrier, and extruded a second time. Part of the core is then replaced with Sn, and this assembly forms what is called a sub-element. After further drawing, the sub-elements are re-stacked and further drawn to the final size. A reaction sequence produces mixed Cu-Sn phases at low to moderate temperatures (below 600°C) before a final high-temperature segment facilitates a diffusion reaction between Cu-Sn and Nb (or Nb-alloy) to form Nb\textsubscript{3}Sn (or Nb\textsubscript{3}Sn alloyed with Ti, Ta, etc.). Conductor naming typically refers to the number of sub-elements that occupy a theoretical number of sub-element sites. RRP research strands explored configurations restacks from 61 up to 217 sub-elements, including Cu.

The tube-based Nb\textsubscript{3}Sn approaches, such as Powder-In-Tube (PIT) and Rod-In-Tube (RIT), replace the annulus of stacked rods in RRP conductors with a solid tube of Nb or Nb-alloy. The Sn source, often combined with Cu, can be a powder or a rod inserted in the center of the tube, whereby a radial diffusion reaction following a similar path as for RRP conductors is used to form the Nb\textsubscript{3}Sn layer. The assembled tube and core material is restacked as described for RRP conductors, where design flexibility is provided by hexagonal or round tube shape, separation and composition of components in the tube core. Since the tube starts off with much less deformation strain than that used to create the RRP annulus, it is plausible that tube-type conductors can achieve larger re-stack counts and smaller diameters before reaching deformation limits. Reports investigated restacks with as many as 744 tube sub-elements (n=18 with 169 sub-elements being Cu) at tube diameter approaching 15 µm for 0.7 mm wire diameter.

There are two areas where tube-like Nb\textsubscript{3}Sn wires need R\&D to improve properties comparable to RRP wires. First, transverse pressure studies of Rutherford cables show that PIT wire tolerate less transverse strain than RRP wires, with respective onset of degradation at 50–120 MPa vs 170–250 MPa. Second, studies of round and rolled wires noted the tendency for PIT reactions to retain a higher degree of radial symmetry than for RRP, which results in a lower conversion of the available Nb to Nb\textsubscript{3}Sn in distorted sub-elements. While high levels of deformation lead to RRR reduction for both designs, the associated loss of Jc for PIT wires presents challenges for magnets. Further enhancement of tube-type Nb\textsubscript{3}Sn conductors could result also from improved manufacturing of tubes with reliable mechanical properties and fine powders of Nb-Sn and Cu-Sn intermetallic compounds.

The continued improvement of Nb\textsubscript{3}Sn wires is vital to future colliders. It includes using advanced Nb-alloys and incorporating high-Cp materials into the wire architecture. Several the potential benefits of using internal oxidation to improve flux pinning and increase J\textsubscript{c} at high field providing retention or increase of B\textsubscript{c2} in Nb\textsubscript{3}Sn conductors is being explored. Development of both RRP and tube routes is presently underway using advanced alloys. These developments are presently bridging from nascent wires to industry R\&D. Adequate conductor development resources could encourage development by industry practical wires over a 5-year time frame. Development of Nb\textsubscript{3}Sn wires with high-Cp additives is underway in parallel with investigation of advanced alloys. Challenges exist in identifying and optimizing powders for both properties and compatibility with manufacturing. Support from conductor development programs could extend promising wires to industry R\&D within 5 years. In addition to these opportunities, improvement of RRP conductors also needs research attention given to a) reducing d\textsubscript{eff} without concomitant loss of performance, loss of yield, or increase of cost; b) scaling production to 100 kg billets or larger; and c) modifying the conductor architecture and heat treatment to optimize properties above 15 T.

\underline{Nb\textsubscript{3}Al composite wires.} The NIMS group in Japan invented a rapid-heating quench treatment (RHQT) process that circumvented thermodynamic limitations of the Nb-Al system. They attained excellent properties of the Nb\textsubscript{3}Al phase in wires of kilometer length. However, the treatment temperature is above the melting point of Cu, which required manufacturing to develop special processes, such as ion plating, to add Cu stabilizer. Conductor development at Ohio State and some cabling and magnet studies at FNAL provide additional background.

\underline{MgB\textsubscript{2} composite wires.} MgB\textsubscript{2}, a binary intermetallic superconductor, is a promising material for application in superconducting magnets. It is brittle, but relatively easy to fabricate in long lengths. The traditional PIT method and a variety of sheath materials with appropriate barriers or reinforcing components are used to fabricate. The MgB\textsubscript{2} superconducting phase is formed during a reaction heat treatment at 700$^o$C for 20 minutes. Since both components Mg and B, and the sheath materials are inexpensive, the cost/performance ratio for MgB\textsubscript{2} is much lower than that for other superconductors. Thanks to the relatively high Tc, MgB\textsubscript{2} is suitable for use in superconducting magnets cooled at 10 to 30 K. The density of MgB\textsubscript{2} is lower than the density of other superconductors making it attractive for specific applications in lightweight magnets or in supporting accelerator magnets as it has been done for the CERN Superconducting Link.
\subsubsection{HTS conductors}
In the 1990s three main HTS materials have been advanced in long length: Bi2223 (Bi\textsubscript{2}Sr\textsubscript{2}Ca\textsubscript{2}Cu\textsubscript{3}O\textsubscript{14}), Bi2212 (Bi\textsubscript{2}Sr\textsubscript{2}CaCu\textsubscript{2}O\textsubscript{10}) and REBCO (REBa\textsubscript{2}Cu\textsubscript{3}O\textsubscript{7}, where RE stands for Y or a rare earth element). Bi2212 round wires and REBCO tapes are the most promising conductors for the use in accelerator magnets due to their capabilities to carry high currents in extremely high magnetic fields.

\underline{Bi2212 round wires.} Bi2212 composites are produced using the PIT method. Round, twisted, multifilamentary Bi2212 wires have significant advantages for application to accelerator magnets. They are fabricated using traditional wire-drawing techniques which permit scale-up and quality control similar to large-scale production of Nb-Ti and Nb\textsubscript{3}Sn wires. Wire twisting efficiently controls conductor magnetization and AC losses.  Standard Rutherford-type cabling techniques allow producing high-current transposed cables. Existing insulation winding and braiding schemes can be adapted with some constraints to accommodate the temperature and chemical interactions for Bi2212. To produce superconducting phase Bi2212 conductor requires final high-temperature reaction  which require using wind-and-react technique for coils made of Bi2212. Present reactions are carried out at 50 bar pressure to achieve the highest superconductor Jc.

Many basic requirements for using Bi2212 in accelerator magnets were addressed over the last two decades by collaboration of the former DOE Office of Electricity program with HEP national laboratories and university groups. Multiple, but small-scale, manufacturers now produce uniform fine powders that facilitate conductor fabrication with low breakage. Continued support from both DOE-HEP and the National Science Foundation (NSF) via the National High Magnetic Field Laboratory (NHMFL) has led to demonstrations of good reproducibility for coil reactions based on conductor witness samples. 

Despite these advances, application of Bi2212 wires has to confront significant challenges due to  high strain sensitivity, high formation reaction temperature in an oxygen-rich environment, and chemical compatibility of the insulation and construction materials during the reaction heat treatment. High intrinsic cost of components and continued availability of conductor from industrial partners are also a challenge for future development and application of this technology.  

\underline{REBCO tapes.} REBCO is produced by deposition on thin tapes that include layers of metals, oxides and ceramics for crystal plane alignment, mechanical strength and electrical stability. The use of high-strength Hastelloy substrates provides mechanical strength and the capability to incorporate pinning centers. However, the Hastelloy thickness should be minimized to provide sufficient  average current density and the ability to form flexible “wires” from REBCO tapes. Conductor on round core (CORC®) and symmetric tape round (STAR®) are two variants envisioned for accelerator magnets. Each consists of REBCO tapes wound in a helical fashion around a conductive core. Thin substrates reduce the strain applied to the REBCO layer, or alternatively permit winding to smaller wire diameter and higher current density.  Wires with typical diameter of ~3 mm and current densities of 1.4 kA/mm$^2$ are made possible by the current state of the art 30 µm thick Hastelloy substrates. However, this development route drives up cost by requiring more conductor per unit length of cored wire and increasing the losses due to damaged material at the slit edges. 

REBCO tapes have been successfully utilized for the 32 T user magnet at NHMFL, Bruker Biospin’s 28.2 T, 1.2 GHz NMR system, and numerous other magnets above 25 T. Several projects are underway aiming at 40 T REBCO based magnets. Significant production currently underway for fusion magnets that provide large amount of data for the evaluation of manufacturing readiness. Smaller production runs associated with the fabrication of large magnets such as the 32 T user magnet at NHMFL have yielded insight about the readiness of this conductor for accelerator magnets. 

\underline{Iron-based superconductors.} A new class of iron-based superconductors (IBS) was discovered in recent years. They show significant potential for high Jc at high fields, and lower cost. Preliminary attempts to fabricate wires and tapes have been made, following either the path of REBCO coated conductors or that of multifilamentary round wires (e.g. powder in tube) but requisites for magnet technology still need to be demonstrated. In particular, addition of substantial high-strength sheathes appear to be necessary to constrain powders and achieve high utilization of the powder core. While continued support of basic research to advance IBS wires is needed, it is unclear where the transition point lies for more focused development for accelerator magnets. High density powder cores must be achieved without requirements of excessive iron or steel sheaths. Coated conductors need to demonstrate advantages over REBCO, and cost obstacles connected with reel-to-reel manufacturing capital must be overcome. 

A practical plan to make the IBS AEFe\textsubscript{2}As\textsubscript{2} (122), where AE = alkaline earth, here Ba or Sr, into a much cheaper, higher stability, round, twisted and effectively isotropic multifilament conductor suitable for 16-20 T magnets is discussed in ~\cite{A7-7}. Whereas the intra-grain Jc is already high enough to compete with Nb\textsubscript{3}Sn, the unpredictable superconducting connectivity at grain boundaries (GBs) is the critical problem that limits the long-range Jc. Obtaining high grain-to-grain Jc is the key breakthrough that would make 122 a practical superconductor with a target Jc above 1500 A/mm$^2$ at 4.2 K and 16-20 T. 

\subsection{High-Current Cables}
Large superconducting accelerator and detector magnets require high-current multistrand cables to limit the piece length requirement for wire manufacturing, the number of turns in coil winding, and the coil inductance. Cables also allow current redistribution between strands in the case of a localized defect or quench. AC losses in cables are controlled by strand transposition and by careful selection of the inter-strand resistance.

\begin{figure}
\begin{center}
\includegraphics[width=0.8\hsize]
{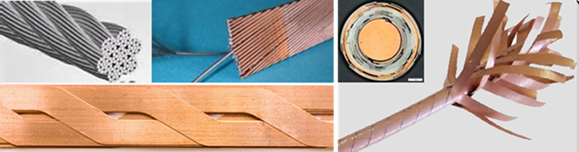}
\end{center}
\caption{Rope-type (top left), Rutherford-type (top middle), Roebel-type (bottom left), and CORC® (right) multistrand superconducting cables. }
\label{fig:AF7-magnets-5.2}
\end{figure}

Four main designs of superconducting cable are shown in  Figure \ref{fig:AF7-magnets-5.2}. They include the rope-type and Rutherford-type designs used with round strands, and the Roebel and CORC® cable suitable for ribbon or tape conductors. Most accelerator magnet designs to date are based on the Rutherford cable concept, using two layers of round wires compressed in a rectangular or slightly trapezoidal shape. This approach is still favored for round conductors, but new concepts are being developed to cope with ribbon or tape conductors.   

The maximum current Ic of a cable depends on strand degradation during cabling and current distribution in the cable cross-section. Due to strand coupling inside the cable, its magnetization and AC losses have additional eddy current components controlled by the inter-strand resistance and strand twist pitch. To reduce coupling, strands can be coated with resistive metal, such as Sn-5\%Ag, Cr, Ni, or a thin resistive core can be used in the cable.

\begin{figure}
\begin{center}
\includegraphics[width=0.8\hsize]
{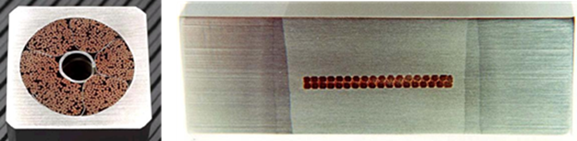}
\end{center}
\caption{Heavily stabilized and mechanically reinforced large current cables: cable-in-conduit-conductor (left) and Al-clad Rutherford cable (right).}
\label{fig:AF7-magnets-5.3}
\end{figure}

In large solenoids used in detectors, additional amounts of a normal metal or stabilizer are used to provide cable mechanical reinforcement in the transverse and axial directions, and improve cable stability and quench protection. Depending on the cooling scheme, gaps or channels for coolant flow can be added. Two examples of such cables are shown in  Figure \ref{fig:AF7-magnets-5.3}. The Cable-In-Conduit-Conductor (CICC) is a version of rope-type cable with strong jacket (conduit) and free space between strands for liquid helium. This arrangement combines good support of the strands against Lorentz forces with a low shielding loss and high current. The copper or Al-clad cables are fabricated using soft-soldering, rolling or co-extrusion processes. High purity Al is used to reduce weight, provide high electric and thermal conductivity, and achieve small radiation length. 
\subsection{Magnet Development and Test Technologies} 
The field range up to 10 T has been scientifically explored and practically realized in all the present HEP colliders such as Tevatron, HERA, RHIC and LHC as well as some smaller machines using Nb-Ti magnet technology. Magnets with a higher field are the most efficient way to achieve higher collision energies in future colliders. To produce higher fields, new superconducting accelerator magnets and technologies based on advanced superconductors and structural materials are being studied. 
\subsubsection{Approaches and Status of high-field Nb\textsubscript{3}Sn accelerator magnet R\&D} 
The present “work horse” that is readily available on an industrial scale is Nb\textsubscript{3}Sn. Over the last three decades, several programs have probed a variety of accelerator magnet concepts, with record dipole fields in multiple configurations ~\cite{A7-54}. As examples:
\begin{itemize}
\item The Cosine-Theta (CT) approach, used in all high-energy collider magnets to-date. The D20 cosine-theta magnet was a flagship 4-layer magnet built in the late 1990’s that achieved a peak field of 13.5 T at 1.9 K in a 50 mm aperture. Most recently, the MDPCT1 dipole magnet, built at FNAL within the US-MDP program, achieved the record fields of 14.1 T at 4.5 K and 14.5 T at 1.9 K in a 60 mm aperture, with excellent field quality ~\cite{A7-55}. 
\item The Common Coil (CC) dipole configuration, wherein vertically-oriented racetrack coils are energized in reverse polarity and create a magnetic field between the coils, yielding effectively twin apertures. Examples include the RD3c magnet built by LBNL, the HFDC01 magnet developed by FNAL, and the DCC017 magnet built and tested by BNL. The latter continues to be in use, serving as a magnet facility for a variety of high field ($\sim$ 10 T) tests for collaborators. 
\item The Block Dipole (BD) magnets, wherein horizontally-oriented racetrack coils energized in same polarity, resulting in a single, high-field bore. To maximize field while allowing access for the particle beam, the ends are flared. The concept was most thoroughly explored by LBNL in the mid 2000’s, culminating in the HD3 magnet that achieved 13.8 T at 4.2 K. The technology was ultimately utilized in the CERN FRESCA2 magnet, which achieved 14.6 T at 4.5 K in a large, $\sim$ 100 mm bore. The magnet was not designed as an accelerator magnet, but rather serves as a test facility for superconducting cables.
\item The Canted Cosine-Theta (CCT) concept, wherein the cable is directly placed in individual grooves machined into a cylindrical support structure. The tilted winding results in a solenoidal field component, which is compensated by the next layer in the structure, and uniform transverse component. The cross-section shows the intrinsic cos-theta current distribution, resulting in excellent field quality.
\end{itemize}
\subsubsection{The 12-14 T operation field range for large scale deployment}
Develop Nb\textsubscript{3}Sn magnet technology for collider-scale production, through robust design, industrial manufacturing processes and cost reduction. The present benchmark for Nb\textsubscript{3}Sn accelerator magnets is HL-LHC, with an ultimate nominal field of 12 T and a production of the order of a few tens of magnets. Nb\textsubscript{3}Sn magnets of this class should be made more robust, considering the full spectrum of electro-thermo-mechanical effects and the processes adapted to an industrial production on the scale of a thousand magnets. The success of this development should be measured through the construction and performance of long demonstrator and prototype magnets, targeting the 12 T field. The Nb\textsubscript{3}Sn magnet development will improve areas of HL-LHC technology that have been found to be sub-optimal, notably the degradation associated with the fragile conductor, targeting the highest practical operating field that can be achieved. The plan is to work jointly with wire and cable development to mitigate degradation associated either with length or electro-thermo-mechanical effects. The R\&D will explore design and technology variants to identify robust design options for the field level targeted. In Europe, the program includes the construction and test of short and long model magnets ~\cite{A7-4}. 

White paper ~\cite{A7-3} discusses the development and demonstration of new technologies for next generation of Nb\textsubscript{3}Sn accelerator magnets with operation fields in the range of 12-14 T. The main goal is to reduce magnet cold-mass cost at least by a factor 2 with respect to the present Nb\textsubscript{3}Sn magnets produced in the US by AUP for the HL-LHC upgrade. To achieve the goal, significant reduction of labour and components costs with higher operating fields and improved performance reproducibility is expected. A key factor for the cost reduction is development of fabrication process based on the experience gained in US national laboratories through the production of the AUP magnets with early industry involvement. This partnership will lead to technologies that will be easily transferable to industry for mid- and large-scale production of Nb\textsubscript{3}Sn accelerator magnets in the 12-14 T field range. This step is critical to enable next generation colliders such as the FCC and Muon Collider. This “directed” R\&D is driven by the field range and industry involvement for high-automation and industry-ready technology. The plan includes 3 phases with total 10 milestones, to be achieved in 6-8 years at the cost of 5-7 \$M/year. This program is proposed as self-standing program, or it may be part of the Leading-Edge technology And Feasibility-directed (LEAF) Program ~\cite{A7-2} aimed at the demonstration of magnet technology readiness for Energy Frontier Circular Colliders (hadron, muon) by the next decade.
\subsubsection{Accelerator magnets in the 14-16 T operation field range}
The projected upper field limit for a Nb\textsubscript{3}Sn dipole is presently 16 T and relies upon mastery of the magnet mechanics. Finding the right balance between cost-efficiency, maximum field, and robustness is at the core of the R\&D activity to provide satisfactory input to the High Energy Physics community. Nb\textsubscript{3}Sn is strain- and stress-sensitive and brittle. Besides the known reversible critical current dependency on applied strain, the main concern is that stress or strain exceeding allowable limits lead to a permanent reduction of critical current and eventual damage through fracture of the superconducting phase. Thus, it is paramount to minimise stress concentrations on the conductor. In the last years, and within the FCC-Conceptual Design Studies, the WP5 of EuroCirCol gathered CEA, CERN, CIEMAT, INFN, KEK, the University of Geneva, the University of Tampere, and the University of Twente to explore different design options for 16 T dipole magnets to give a baseline for future development. The focus was on the design of electromagnetically efficient coil cross sections. In the US, the US-MDP program focused in the stress management approaches to reduce the stress in the strain-sensitive Nb\textsubscript{3}Sn conductor.

Advanced magnet technology is the driving technology for any energy-frontier circular collider envisioned by the community, fundamentally impacting both science reach and facility cost. There are many elements of technology development that lie at the heart of advanced magnet R\&D. It includes new diagnostics and testing techniques, materials and composites development, new solutions for detection and protection and the development of modelling tools together with the required specialized testing infrastructure. Finally, dedicated manufacturing and test infrastructure required, including instrumentation upgrades, needs to be developed, built and operated through close coordination between the participating laboratories.  
\subsubsection{Coil stress management}
\begin{figure}
\begin{center}
\includegraphics[width=0.6\hsize]
{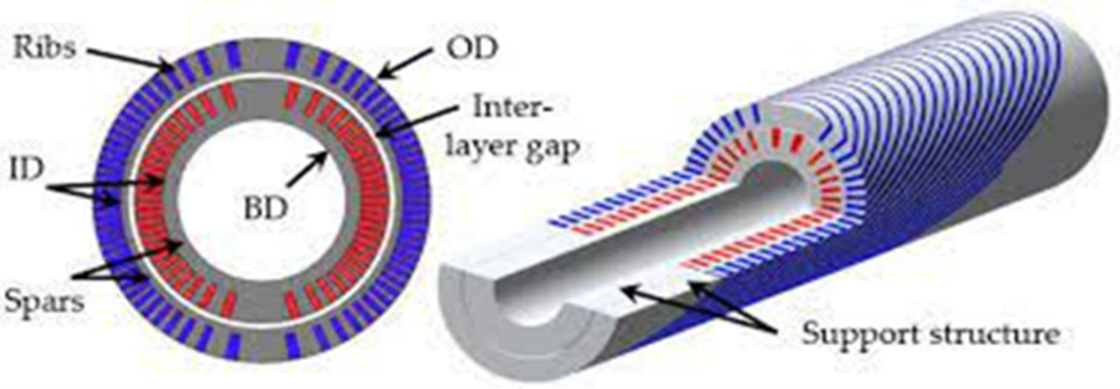}
\includegraphics[width=0.6\hsize]
{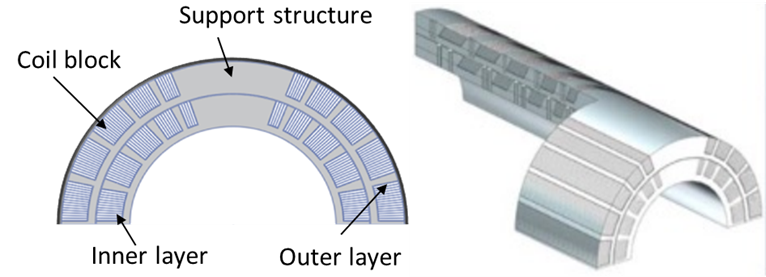}
\end{center}
\caption{Top - the Canted Cosine Theta (CCT) concept, wherein the individual turns are placed and supported in support structure grooves. Bottom - the Stress Managed Cosine Theta (SMCT) concept, wherein turn groups are placed and supported in support structure grooves.}
\label{fig:AF7-magnets-5.4}
\end{figure}

To reduce conductor stress, the US MDP program ~\cite{A7-1} is focused on stress-management approaches introduce internal structures and additional interfaces when compared to more traditional magnet designs. The concept of stress-management, i.e., intercepting magnetic forces to structural members before they can accumulate to damaging levels, promises a path towards very high field accelerator magnets. Two stress management design approaches are shown in  Figure \ref{fig:AF7-magnets-5.4}.

To take full advantage of these technologies, further understanding and development in several areas is necessary. Below are some key issues that are being investigated to develop these stress-management approaches: 
\begin{itemize}
\item Development stress management coil structures and their fabrication methods including exploration of conventional machining methods as well as additive manufacturing approaches 
\item Investigate and optimize external structures for use in stress-managed magnets with increases internal structure rigidity 
\item Understand the influence of interfaces on stress-limits and training under different interface conditions 
\item Explore the limits of stress-management approaches on high field / large bore demonstrators
\item Develop efficient and cost-effective fabrication methods using minimal tooling to reduce overall magnet fabrication cost and complexity 
\item Investigate approaches for scale up to larger magnets
\end{itemize}
\subsubsection{Considerations for accelerator magnets beyond 20 T}
Bore fields at 15 to 16 T level are considered as the practical limit for Nb\textsubscript{3}Sn superconductor ~\cite{A7-54}. To further push the magnetic field of dipole magnets, HTS need to be considered. Taking into account the higher HTS cost and the present level of HTS magnet technologies with respect to the Nb\textsubscript{3}Sn conductors and magnets, hybrid approach to the magnet design looks the most practical. 

\begin{figure}
\begin{center}
\includegraphics[width=1.0\hsize]
{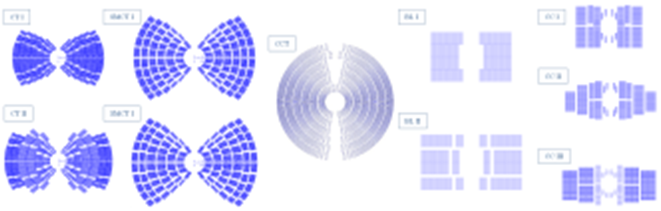}
\end{center}
\caption{Preliminary cross-sections of 20 T hybrid dipole coils. From left two right: Cos-theta (CT) design, with 4 (top) and 2 (bottom) layer Bi2212 coils; Stress management Cos-theta (SMCT) design, with 4 (top) and 2 (bottom) layers Bi2212 coils; Canted Cos-theta (CCT) design, with 4-layer Bi2212 coil; Block (BL) design, with and without stress management; Common Coil (CC) design, with Bi2212 (top, with 3 external Nb\textsubscript{3}Sn layers, and center, with 5 external Nb\textsubscript{3}Sn layers) and REBCO CORC® coils (bottom, with 4 external Nb\textsubscript{3}Sn layers). For all the CC designs, only one aperture is shown.}
\label{fig:AF7-magnets-5.5}
\end{figure}

Judicious magnet design, incorporating concepts such as stress management and optimized pre-stressing, indicates 20 T dipole fields may be achievable with existing superconducting materials in a variety of magnet configurations. Magnetic analysis of a 20 T hybrid magnet in a 50 mm aperture with at least 15\% of load-line margin are on-going within MDP (see ~\cite{A7-1} and ~\cite{A7-12}), considering different layouts: traditional Cos-theta (CT), Stress Management Cos-theta (SMCT), Canted Cos-theta (CCT), Block (BC), and Common-Coil (CC) designs (see   Figure \ref{fig:AF7-magnets-5.5}). Two HTS conductors are considered at the present time: Bi2212, in the form of a Rutherford cable with Je of 740 A/mm$^2$ at 20 T, and REBCO tape in a CORC®/STAR® cable with Je of 590 A/mm$^2$ at 20 T. 

\begin{figure}
\begin{center}
\includegraphics[width=0.6\hsize]
{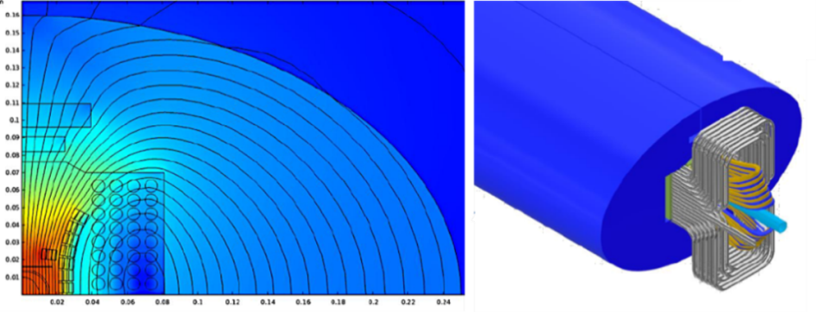}
\end{center}
\caption{TAMU Hybrid dipole models.}
\label{fig:AF7-magnets-5.6}
\end{figure}

A strategy by which a tape-stack REBCO cable may be configured in a conformal winding in such a way that the favourable parallel orientation is sustained everywhere in the winding is presented and discussed in ~\cite{A7-14}. A dipole geometry is presented in which such a conformal REBCO insert winding is configured within a cable-in-conduit Nb\textsubscript{3}Sn outsert winding so that the sub-windings can be separately fabricated, and then assembled in a common structure and preloaded for effective stress management. The total cost of superconductor in this dipole is a minimum among the design options for 18 T. Plans for development of the coil technologies and a model dipole are presented.  Figure \ref{fig:AF7-magnets-5.6} shows an example field design for such a dipole. The insert winding consists of a tape-stack cable in which 25 REBCO tapes are stacked face-to-face, and each turn of the cable is oriented closely parallel to the field at that location.

\subsubsection{HTS insert technology R\&D plan}
The HTS inserts are the key most complicated part of the hybrid magnet design. The plans to develop HTS dipole inserts for high field hybrid magnets are discussed below based on submitted WPs. Technologies of Nb\textsubscript{3}Sn outserts will be produced as a part of Nb\textsubscript{3}Sn HFM R\&D.

\underline{Bi2212 inserts}. Bi2212 model magnet R\&D program is discussed in ~\cite{A7-13}. As the only high-temperature superconductor (HTS) available as an isotropic, twisted, multifilamentary round wire, Bi2212 is very promising for expanding the high-field superconducting accelerator magnet technologies beyond the Nb-Ti and Nb\textsubscript{3}Sn. The document describes the technology status and define appropriate steps to develop Bi2212 into a practical very high field (above 15 T for accelerator magnets, and above 25 T for solenoids) magnet technology, from superconductor development, to design and development leading to the construction of model coils and magnets fulfilling accelerator-quality needs for muon or hadron colliders. The first and most relevant target is to develop high-field accelerator dipole magnets capable of generating 16-20 T in a bore of 40-50 mm suitable for colliders. The second target is to develop large-bore, high-field 15+ T quadrupole magnets for interaction regions. A third target is to develop 25 T or greater commercial solenoid magnets in collaboration with the US magnet industry and to develop the essential elements of the technology needed to build 30+ T solenoids for muon colliders. Given the potential helium shortage, a fourth relevant target is to develop Bi2212 magnets, both solenoids and accelerator magnets, for other special uses at higher operation temperatures (10-25 K). 

There are certain challenges to be studied and addressed for Bi2212 magnet technology. Preliminary data show that Bi2212 Rutherford cables can show 5\% critical current degradation under $\sim$ 120 MPa transverse pressure. To increase the safety margin, two stress management accelerator magnet concepts have been proposed, the Canted Cosine-Theta (CCT) at LBNL and the Stress Management Cosine-Theta (SMCT) at FNAL. 

Similar to Nb\textsubscript{3}Sn, Bi2212 magnets are fabricated using a wind and react process. A key advance for Bi2212 since the 2014 P5 is the introduction of the overpressure processing heat treatment, which uses a high-pressure mixed Argon and Oxygen (50 bar typically with 98\% Ar and 2\% O\textsubscript{2}) to densify the superconducting filaments and obtain a high wire Je. For single wires a TiO\textsubscript{2} coat on the wire and a braided mullite insulation works well but this is not yet working as well for Rutherford cables because an application of a robust TiO\textsubscript{2} has not yet been achieved. The result is some Bi2212 leakage through the encasing metal (Ag or Ag alloy) at high temperatures which can react with surrounding SiO\textsubscript{2}-containing insulation. Leakage degrades wire Je because it depletes a portion of filaments and interrupts current flow. Methods to evade these restrictions are under active development.

Nb-Ti and Nb\textsubscript{3}Sn magnets are prone to quench due to a low Tc and enthalpy margin. However, it is much easier to spread out the stored energy for Nb-Ti and Nb\textsubscript{3}Sn magnets due to fast normal zone propagation velocity. Slow growth of normal zones in HTS magnets makes reliable quench detection critical to reliable quench protection.

To fabricate accelerator magnets of 15+ T and solenoids of 25+ T, HTS coils are combined with Nb-Ti and Nb\textsubscript{3}Sn to minimize overall costs. Typical designs use HTS inserts nested concentrically inside Nb-Ti and/or Nb\textsubscript{3}Sn outsert magnets, setting up issues of magnet assembly and integration for both accelerator and solenoid magnets given their very different stabilities and quench velocities. This magnet integration will require careful modeling, design, and experimental validation.

The length of main ring dipole magnets for circular colliders ranges within 5-15 m. Scaling the magnet fabrication from $\sim$ 1 m long models up to 5-15 m full-scale magnets presents another challenge. The coil reaction will require building heat treatment facilities adequate to handle such long magnets. The magnet stored energy goes up linearly with the magnet length and, thus, quench protection may become an issue.

Proposed paths to developing Bi2212 magnet technologies for accelerator applications include:
\begin{itemize}
    \item All Bi2212 magnets depend on a stable supply of good conductors with reliable properties. As for Nb-Ti and Nb\textsubscript{3}Sn, an essential component of the background R\&D is to understand the conductor, especially present limitations on properties so that ongoing collaborations in the university-industry-lab nexus can drive further improvements.
    \item Use CCT and SMCT Bi2212 designs as R\&D vehicles to drive conductor development and accelerator magnet design, technology, and test towards maturity. The goal is to build multiple short (up to 1 m long) dipole model magnets with a bore of 40-50 mm in standalone configuration with incremental dipole field from 5 T to 10 T. Several critical technology elements are to be optimized and developed: a) Master technology know-how of the reaction of $\sim$ 1 m long coils and demonstrate reliability. b) Optimize Bi2212 Rutherford cable fabrication including its packing factor and insulation, especially to minimize the contribution of insulation materials to the thermodynamic ceramic leakage.
    In hybrid magnet configuration, fully explore the hybrid magnet technology and its reproducibility. Gain experience with magnet operation and quench protection for the new kind hybrid magnets.
    With technology maturing, build multiple, $\sim$ 1 m long 16+ T Nb\textsubscript{3}Sn/Bi2212 hybrid accelerator dipole magnets with a projected bore diameter of 40-50 mm. Gain fabrication and operation experience.
    \item Perform a critical assessment in the feasibility of scaling up accelerator magnets (technologies and infrastructure, etc.) to 5-15 m long.
    \item For both accelerator and solenoid magnets, explore special uses at higher operation temperatures (10-25 K).
    \item Engage the US superconductor and magnet industry to develop production capabilities for large scale raw materials, wire and cable fabrication, and to reduce cost and supply chain risk.
\end{itemize}

The Bi2212 dipole insert program started recently at FNAL as part of the US-MDP is discussed in ~\cite{A7-21}.  The goal of this program is to develop and test the technology of HTS inserts based on Bi2212 cable and cos-theta coil configuration. With existing Bi2212 superconducting strands and cables, the potential reach for the maximum field is of up to 19 T, thanks to the progress realized in wires. To achieve Bi2212 potential in coils, however, a number of technological questions still have to be answered, including the stress limit for Bi2212 magnet design, insulation processing and materials that prevent leaks which reduce Bi2212 conductor transport current, and if needed, optimization of conductor surface in contact with Oxygen for coils made of Rutherford cable. The presented insert design concept, as well as the basic technological solutions, will be studied experimentally and optimized on a series of short coils and models.

\underline{REBCO inserts.} REBCO model magnet R\&D plan is discussed in ~\cite{A7-11}. The paper concerns two issues: the ultimate high field to meet the physics needs, and the ultimate low cost. REBCO material has a high irreversibility field with an upper limit of 110 T at 4.2 K and 100 T at 20 K. Such a high irreversibility can allow REBCO magnets to generate a dipole field of at least 20 T over a temperature range of 1.9-20 K. REBCO coated conductors have significant room for cost reduction due to the raw material cost. Operating at elevated temperatures without liquid Helium can be another opportunity that REBCO can offer to further reduce the magnet and collider operation cost.

Although REBCO shows great potential for addressing the high-field and low-cost needs, the challenges to realize the potential are significant for two reasons. First, REBCO magnet and conductor R\&D is at the very yearly stage. Second, more conductors are needed to make magnets and to learn in a sufficiently fast pace. The following steps are recommended to improve the situation:
\begin{itemize}
    \item Engage REBCO conductor vendors and couple the development of conductor and magnet technologies. Use the magnet results as critical feedback to the conductor development that, in turn, can help improve the magnet performance.
    \item When sufficient funding is available, fast-track the development with the goal to quickly progress on the maximum dipole field a REBCO magnet can generate. Aim for 10 T dipole field within three years and 15 T within five years.
    \item Collaborate with the fusion magnet community and support the development of REBCO fusion magnet systems.
\end{itemize}

Two material properties of REBCO exacerbate the situation. 

First, the ceramic REBCO material is brittle. The REBCO layer can crack, when subject to a tensile strain of around 0.6\% and higher, and permanently degrade its current-carrying capability. Bending or twisting the tapes that are necessary to wind coils or making multi-tape cable must not exceed this strain limit. REBCO coated conductors are also weak to withstand a tensile force applied transverse to its broad surface; a stress of several MPa can delaminate the tape and degrade the REBCO layer. Epoxy impregnation can degrade REBCO conductor if the thermal contraction of the epoxy mismatches that of the conductor. Similar mechanical issues appear again during magnet operation when the Lorentz forces become excessive on the conductor. The maximum dipole field a REBCO magnet can generate will likely be determined by the mechanical limit of the conductor. One particular concern from the brittleness of REBCO is the potential performance degradation in conductor and magnet. Such a degradation has been observed in high-current REBCO fusion cable samples.

Second, REBCO coated conductors are only available as a tape with an aspect ratio of at least 10. The tape conductor can be bent, but only as a developable surface with limited flexibility. This geometric constraint limits the potential magnet designs that one can work with REBCO conductors. One solution is a round-wire conductor form assembled from multiple tapes, such as CORC® and STAR® cables, both are actively pursued in the U.S. with a strong support from SBIR programs.

High-field dipole magnets require large-current conductors, operating at 10-20 kA, to reduce the magnet inductance and to accelerate current ramping. Various concepts exist for multi-tape REBCO cable for high-field magnets, such as twisted-tape stack and Roebel, CORC® and STAR® cables. CERN developed dipole magnets using a stack of tapes and Roebel cables. The U.S. MDP is developing dipole magnets using CORC® cables with different magnet concepts. The current maximum dipole field achieved by a REBCO magnet is 5.4 T at 4.2 K in a racetrack magnet from the European EuCARD program and 4.5 T at 4.2 K in a Roebel-cable based magnet from the European EuCARD2 program.
There is a significant technology gap between where we are today and a REBCO high-field dipole magnet. Here are some important questions that need to be addressed for REBCO dipole magnet and conductor technologies:
\begin{itemize}
    \item How to make high-field accelerator magnets using multi-tape REBCO conductor?
    \item What is the maximum field a REBCO dipole magnet can achieve? What factors limit the maximum dipole field of a REBCO magnet and how can we address them? How to develop magnet structures to limit stress on conductors? What is the long-term performance of REBCO magnets under Lorentz loads? 
    \item How do REBCO magnets transition from superconducting to normal state and how can we detect the transition? 
    \item What is the field quality of REBCO accelerator magnets?
    \item What is the required performance for REBCO conductors to achieve the desired magnet performance?
    \item How to determine the performance of a long multi-tape REBCO conductor for predictable magnet performance? 
\end{itemize} 
\subsubsection{Diagnostics and testing techniques, technology, advanced modeling.} 
New diagnostics and testing techniques are providing important insight into magnet behaviour and provide critical feedback to conductor and magnet designers. The main diagnostics and testing technique development plans are discussed in, ~\cite{A7-15}-~\cite{A7-18}. They include:

\begin{itemize}
    \item Develop a next-generation acoustic emission diagnostic hardware capable of self-calibration to improve disturbance triangulation accuracy and “fingerprinting”;
    \item Establish fiber-optic based diagnostic capabilities through the use of Fiber Bragg Grating (FBG) and Rayleigh scattering-based sensors
    \item Develop new methods for reliable and robust quench detection and localization for HTS magnets and hybrid LTS/HTS magnets using Rayleigh sensors
    \item Develop flexible multi-element quench antennas, large-scale Hall sensor arrays and non-rotating field quality probes, aiming at understanding electromagnetic instabilities in LTS magnets and imaging current-sharing patterns in superconducting cables and HTS magnet coils
    \item Develop new algorithms for current flow reconstruction and disturbance localizations
    \item Apply machine learning and deep learning approaches to process diagnostic data and identify real-time predictors of magnet quenching
    \item Develop non-invasive surface and insulation failure detection and localization using electrical reflectometry, to protect magnets due to unexpected defects on the surface or insulation
    \item Develop cryogenic digital and analog electronics to facilitate, simplify and improve reliability of diagnostic instrumentation by enabling pre-processing of magnet diagnostic data in the cryogenic environment
\end{itemize}

A number of potential techniques and technologies aiming to affect magnet parameters and performance need to be further investigated including a) the development of high-Cp conductor and insulation that can lead to conductors and coils with optimized characteristics enabling stable operation against perturbations; b) artificially increasing the coil current during a quench by discharging a large capacitor at quench detection; c) using diffuse field ultrasonic techniques to enable targeted delivery of vibrational excitation to the conductor, for a non-invasive structural local probing of SC coils; d) development of composite and structural magnet components, with a specific focus on the characterisation of insulation systems (polymers and reinforcement) for both Nb\textsubscript{3}Sn and HTS magnets; e) thermal management of high field magnets including internal heat transfer, heat transfer to coolant, and external heat transfer to cryo-plant.

Advances in modelling are providing more detailed insight into stress and strain states in magnets, including full 3D effects and complex nonlinear material and interface behaviour. An analysis of the data acquired by the diverse diagnostic tools and advance modelling is the key to guide the potential techniques to affect magnet performance. A diverse and challenging set of new modelling tools are required to continue this effort and ultimately improve design time, cost, and performance of future superconducting accelerator magnets. The new developments include simulation and advanced modelling of a) conductors and cables, b) interfaces and other potential sources of training in stress-managed designs, c) LTS/HTS hybrid magnets, and d) radiation-induced thermal effects on magnets in synergies with the broader accelerator modelling community ~\cite{A7-22}.
\subsubsection{Magnet Fabrication and Test infrastructures} 
The development of novel SC magnet technology at the high-field frontier requires specialised fabrication and test infrastructure, often of large size. The necessary investment is considerable, so the effort should extend across national and international programs and collaborations. In the U.S. the magnet R\&D infrastructure exists at the three National Laboratories (BNL, FNAL, LBNL) as well as at NHMFL and TAMU. It includes R\&D magnet development and test as well as small production capabilities. Some of these capabilities are unique. BNL and FNAL developed and operate short and long coil winding-reaction-impregnation facilities as well as short and long magnet assembly and test facilities. This infrastructure can be and is also used for production of small quantities of full-scale accelerator magnets, e.g., for the HL-LHC upgrade. LBNL has largest in the world 60-strand cabling machine which provides Rutherford cables for various accelerator magnet R\&D programs and production. BNL offers 10 T common coil facility for R\&D cable and insert magnet testing. NHMFL developed and built exclusive ovens for HTS coil reaction at high pressures. 

The work on the infrastructure continues. FNAL in collaboration with LBNL is building a new high-field cable-testing facility ~\cite{A7-56} with capabilities similar to the European EDIPO and FRESCA2. This facility will serve two U.S. national programs within the DOE Office of Science – the US Magnet Development Program and the US Fusion Energy Science Program – for testing HTS samples in magnetic fields up to 15 T. This facility will allow testing HTS-LTS hybrid magnets which are an important step toward providing 20+ T operation fields in future high-energy colliders. This new test facility will be available at FNAL within the next 3-5 years.

All these facilities need adequate support to operate, maintain and upgrade. Opportunities to reduce power requirements for facility operation support for test facility upgrades and operation need to be explored and used. Wide collaboration with various partners inside the U.S. and with laboratories, universities and industry in Europe and Japan on magnet material and component development and test is critical to achieve results on time with minimal expenses.
\subsubsection{Technologies for Detector Magnets}
The development of technologies for detector magnets is discussed in ~\cite{A7-19}. The main development item for future detector solenoids is Al-stabilized superconducting cable with both higher strength and high RRR. The most likely solution is a combination of technologies used in ATLAS-CS (reinforcement of the aluminum stabilizer, with Ni doping and simultaneous cold-work hardening) and CMS (reinforcement by pure-aluminum stabilized conductor and high strength aluminum alloy, which are mechanically bonded by electron beam welding). By combining these approaches, a yield strength of more than 300 MPa is expected. Future detector magnets will also take advantage of technologies such as coil winding inside the outer shell; indirect cooling to reduce materials used in magnet structure; pure aluminum strips as temperature equalizer in the steady operation and fast quench propagation; lightweight and high-transparency vacuum vessels. High temperature superconductors are of great interest to reduce cryogenics power consumption of detector magnets. While solenoids have been prevalent in past detectors, other types of magnets, such as split coil and saddle shape coil, could be candidates depending on the requirements from a physics viewpoint. Solenoids for mid-scale experiments often do not need the development of unique conductors, as much as precise control of magnetic field distribution with precision simulation and coil fabrication technology to meet tight tolerances.

\section{Conclusions}
Superconducting accelerator magnets are the key enabling technology for present and future particle accelerators in modern high energy physics. All the present accelerator magnets have used for decades the Nb-Ti superconductor. The practical performance limit of this technology in accelerator magnets is limited by 8-9 T. This field level was realized in various LHC magnets. The technology will continue to be developed and widely used in various accelerators, such as ILC, FCC-ee, EIC, SEPC, etc., to steer beams of low-energy or small-mass particle. This technology continues to be the baseline approach for large detector magnets. 

The development of a new generation of accelerator magnets based on Nb\textsubscript{3}Sn superconductor with operation fields up to 15-16 T has shown in the last two decades a good progress in the USA, EU and Asia. On the short term, within the next five years, this technology will be implemented in HL-LHC by using large-aperture high-gradient final-focus quadrupoles in ATLAS and CMS experiments. On the longer term, dipole and quadrupole magnets with nominal operation fields up to 16 T are planned for FCC-hh and 3 TeV MC. Several key technological issues important for the performance of these magnets need to be resolved. They are in the focus of national magnet R\&D programs in USA and EU. Demonstrating performance retention throughout the whole magnet life, mastering electro- and thermo-mechanical loads, and achieving a considerable simplification of manufacturing will be necessary to exploit the Nb\textsubscript{3}Sn magnet technology on large scale.

More ambitious R\&D work towards 20+ T magnets, considered for SPPC and high-energy MC and based on cost-effective HTS/LTS coils has also been started recently. HTS materials promise higher fields and improved energy efficiency through operation at higher cryogenic temperature, or helium free configurations. This technology is still at its early stage. Much work needs to be done on the way to realize the potential of HTS materials, starting from their basic conductor characteristics through cable and magnet design, technology and quench protection, including their mass production and cost optimization issues.

\end{document}